\documentclass[12pt]{article}
\usepackage{float}
\usepackage[margin=1in]{geometry}
\usepackage{amsmath,amssymb,bm}
\usepackage{graphicx}
\usepackage{booktabs}
\usepackage{array}
\usepackage{authblk}
\usepackage[numbers,sort&compress]{natbib}
\usepackage[hidelinks]{hyperref}
\usepackage[nameinlink]{cleveref}
\usepackage{xcolor}
\usepackage{caption}

\newcommand{\dd}{\mathrm{d}}

\title{Tunable inertia controls rigid-body flipping in microgravity}

\author[1]{Briggs Pugner}
\author[1]{Justin Derickson}
\author[2]{William L. Weaver}
\author[1,3,*]{Aaron H. Persad}
\author[2,*]{Kausik S. Das}

\affil[1]{Department of Engineering and Aviation Sciences, University of Maryland Eastern Shore, Princess Anne, Maryland 21853, USA}
\affil[2]{Department of Natural Sciences, University of Maryland Eastern Shore, Princess Anne, Maryland 21853, USA}
\affil[3]{International Institute for Astronautical Sciences}
\affil[*]{Corresponding authors: ahpersad@umes.edu; kdas@umes.edu}

\date{}

\begin{document}
\maketitle

\begin{abstract}
Rotational stability in microgravity depends critically on the distribution of mass within a freely rotating body. Here we demonstrate that intermediate-axis instability, known as the Dzhanibekov effect, can be enabled or suppressed by tuning payload geometry. Independent inertia measurements and a CAD-based digital twin predict an instability window between flight-mount angles of approximately $9.5^\circ$ and $54.5^\circ$. Parabolic-flight experiments confirm the predicted stability classification: repeated inversions occur at $30^\circ$ and $42.9^\circ$, whereas no flipping is observed at $0^\circ$, $60^\circ$, and $75^\circ$. Onboard inertial measurements are consistent with torque-free Euler dynamics, with a measured inversion interval of $0.420\pm0.007$ s compared with a numerical prediction of $0.421$ s. Poinsot reconstruction and an elliptic-integral solution connect the observed inversions to nonlinear separatrix dynamics. These results establish a microgravity-tested framework for predicting geometry-dependent rotational stability in spacecraft, deployable structures, and free-flying robotic systems.
\end{abstract}


\section{Introduction}

Rotational stability is a central challenge for objects moving in microgravity, from satellites and deployable spacecraft to tumbling space debris, free-flying robots, rotating space habitats, and natural bodies with irregular mass distributions \citep{Dipert2023}. In such systems, attitude motion is not determined by angular momentum conservation alone. It is also determined by the geometry of the inertia tensor. Small changes in mass distribution can reorganize the qualitative structure of torque-free motion, moving a body from steady spin to large-amplitude reorientation even in the absence of external torque.

A canonical example of this sensitivity is the intermediate-axis instability, commonly known as the Dzhanibekov effect or tennis-racket theorem\citep{EulerRigidBody} - \citep{HughesSpacecraft}. A rigid body rotating about the principal axis with the largest or smallest moment of inertia is linearly stable, whereas rotation about the intermediate principal axis is unstable. The resulting motion can appear paradoxical: angular momentum and rotational kinetic energy remain conserved, yet the body undergoes repeated inversions. These flips are not caused by a hidden impulse or unmodeled torque. They arise from the intrinsic geometry of torque-free rotation.

The same principle appears across many physical systems. In spacecraft engineering, changes in mass distribution caused by deployable structures, robotic appendages, fuel motion, or variable-geometry components can alter attitude stability and move a vehicle into or out of a tumbling regime. Natural bodies such as asteroids and space debris exhibit complex weakly torqued or nearly torque-free rotations whose long-time behavior depends on the ordering of their principal moments of inertia. In human and animal motion, high divers, gymnasts, and falling animals use controlled changes in body shape to modify the inertia tensor and produce rapid reorientation. These examples show that intermediate-axis dynamics are not an isolated curiosity, but a representative mechanism by which geometry, mass distribution, and conserved angular momentum produce nonlinear attitude motion.

Despite the long theoretical history of torque-free rigid-body dynamics, quantitative experimental mapping of the transition into and out of intermediate-axis instability remains limited. Familiar demonstrations use fixed objects such as books, wing nuts, tennis rackets, smartphones, or hand tools. These demonstrations show that the instability exists, but they do not reveal how the instability appears and disappears as the inertia tensor is tuned. Ground-based measurements are further complicated by gravitational torques, supports, bearings, friction, aerodynamic drag, and release imperfections. Microgravity provides a cleaner environment for approximating torque-free motion, yet most microgravity demonstrations observe the effect for one mass distribution rather than mapping the stability transition of a controlled variable-inertia body.

Here we introduce a tunable rigid-body platform that maps intermediate-axis instability in microgravity. The payload uses adjustable radial arms to vary the principal moments of inertia while preserving a well-defined body frame. By changing the arm angle, the same physical object can be moved from stable rotation, through an intermediate-axis instability window, and back to stable rotation. The experiment therefore turns the Dzhanibekov effect from a binary demonstration into a controllable transition in inertia space.

\begin{figure}[t]
\centering
\includegraphics[width=0.8\linewidth]{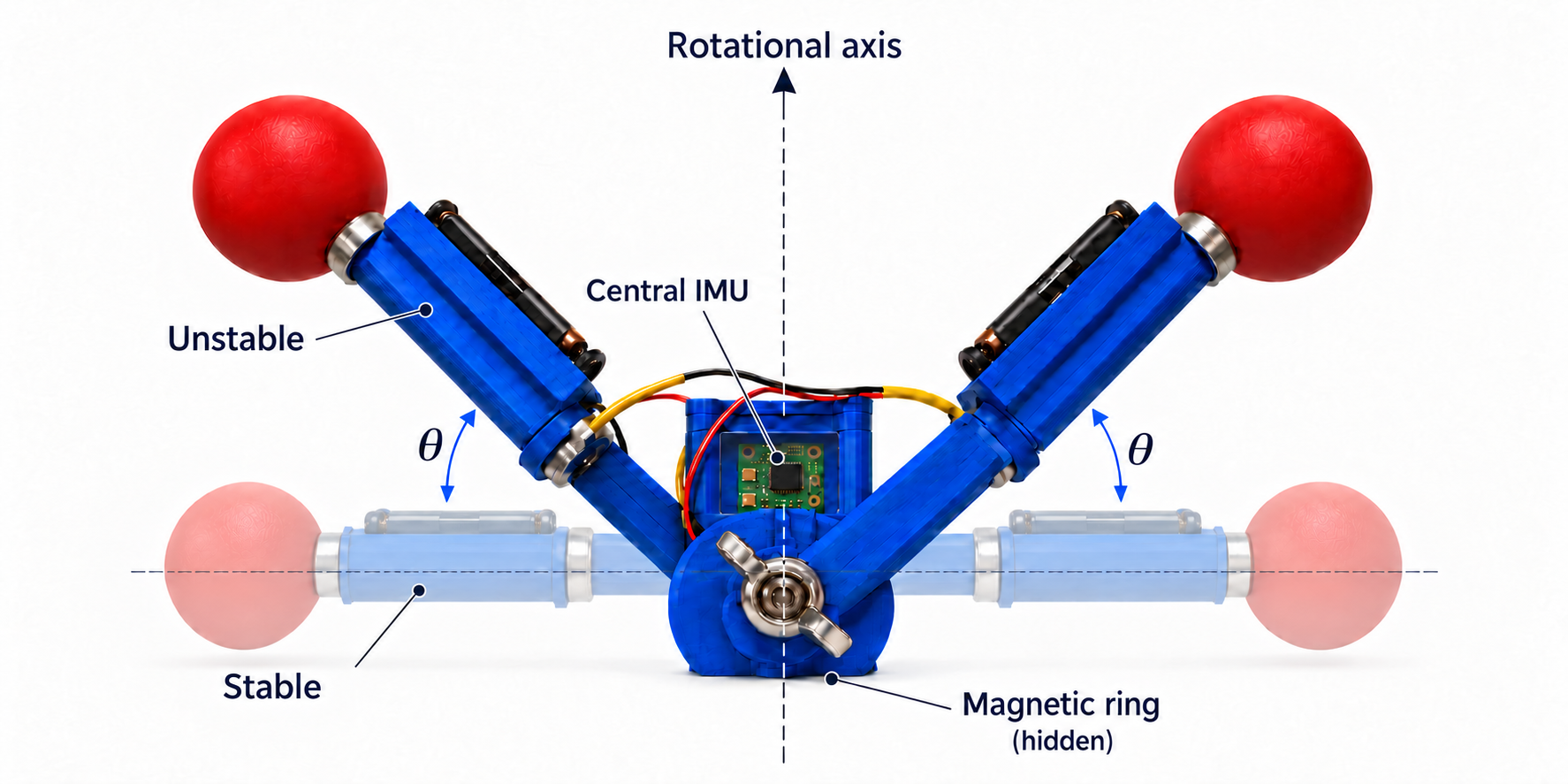}
\caption{
 Schematic of the tunable rigid-body payload, adjustable arms, arm angle $\theta$, body-fixed axes, spin-up direction, electromagnet release fixture, parabolic-flight microgravity interval, and onboard IMU. This object is tuned by varying $\theta$ through stable and unstable inertia configurations, rotated about the rotational axis and then released in microgravity.}
\label{fig:V_shape}
\end{figure}

The present study integrates experimental measurements, computational
modeling, and nonlinear rigid-body theory to investigate the
geometry-dependent onset of intermediate-axis instability in
microgravity. First, a tunable rigid-body payload was designed
and constructed with adjustable arms, allowing systematic
modification of its mass distribution and principal moments
of inertia (Fig. \ref{fig:V_shape}). The principal moments of inertia were independently
measured on the ground using a falling-mass apparatus, with
center-of-mass measurements and parallel-axis corrections
applied where necessary (Fig. \ref{fig:ground_measurement}). A CAD-based digital twin was then
developed to calculate the principal moments of inertia as
functions of the flight-mount angle $\theta$. Comparison
with the ground-based measurements validated the computational
model and enabled identification of the angular configurations
corresponding to stable and unstable rotation about the
nominal spin axis.

Guided by the predicted stability map, selected payload
configurations were tested during parabolic flights. During
each microgravity interval, the payload was spun to a
controlled angular speed and released, while an onboard
inertial measurement unit (IMU) recorded the three
components of angular velocity. The experimentally observed
rotational dynamics were subsequently compared with
numerical solutions of the torque-free Euler equations
using the measured principal moments of inertia.
Poinsot's geometric construction was employed to
interpret the trajectories in angular momentum space,
while nonlinear rigid-body theory and elliptic-integral
solutions provided predictions for the characteristic
flip times and their dependence on the inertia ratios.

By combining independently validated inertia measurements
with controlled microgravity experiments and theoretical
predictions, this approach directly connects the geometry
of a rigid body to its rotational stability and nonlinear
dynamics. In particular, it provides a framework for
identifying the onset and suppression of intermediate-axis
flipping, characterizing the resulting angular-velocity
evolution, and examining the relationship between the
principal moments of inertia and the observed flip times.

\section{Results}

\subsection{Ground Measurement and Digital Twin Validation of
the Principal Moments of Inertia}

\begin{figure}[htbp]
\centering
\includegraphics[width=0.7\linewidth]{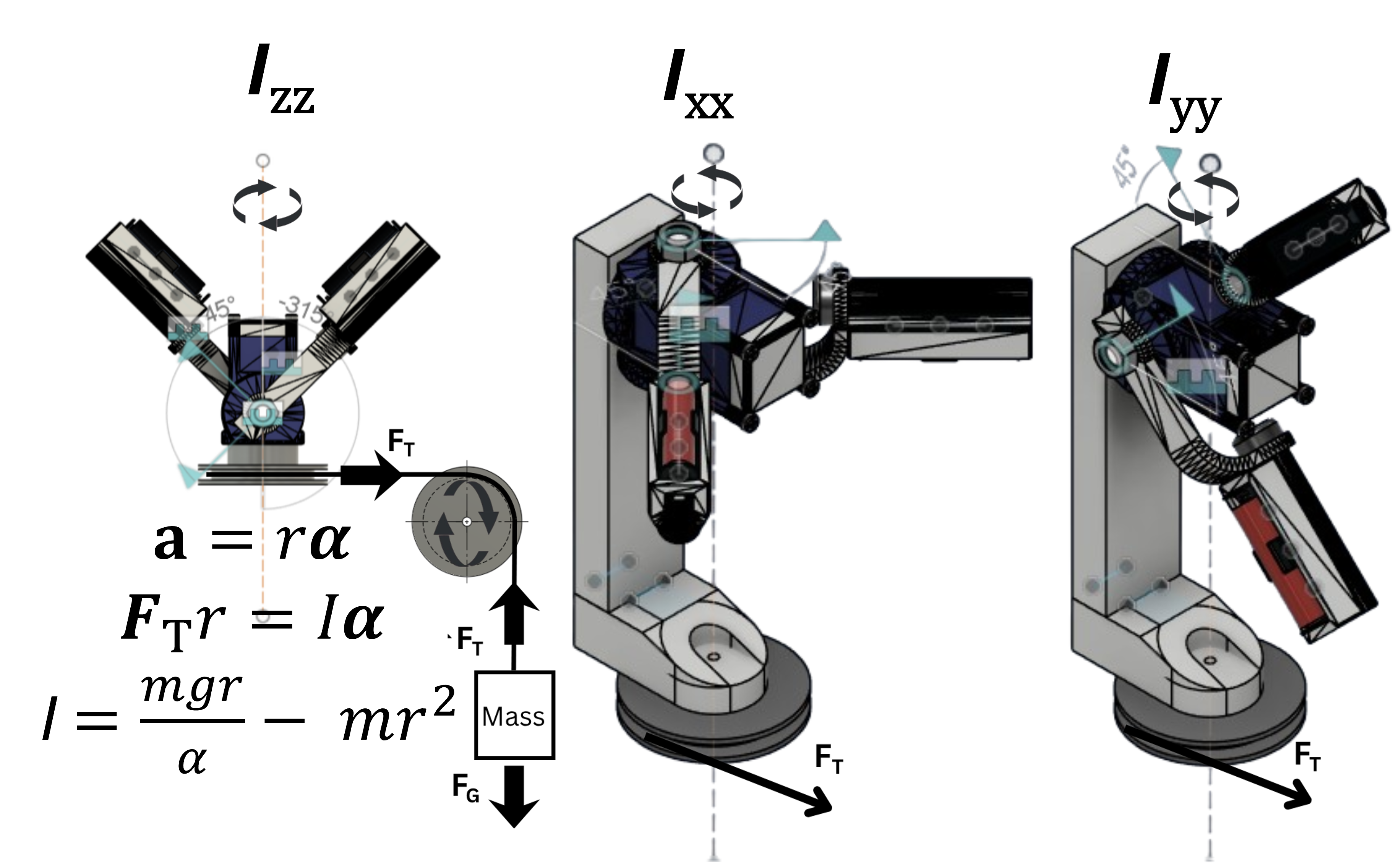}
\caption{
\textbf{Ground-based measurement of the principal
moments of inertia.}
Experimental pulley and falling-mass apparatus used
to determine $I_{xx}$, $I_{yy}$, and $I_{zz}$ of the
payload. Measurements were corrected for the
position of the center of mass using the
parallel-axis theorem where necessary.
}
\label{fig:ground_measurement}
\end{figure}

The principal moments of inertia, $I_{xx}$, $I_{yy}$,
and $I_{zz}$, of the experimental payload were independently
measured on the ground using a pulley and falling-mass
apparatus, as illustrated in Fig. \ref{fig:ground_measurement}.
The payload was mounted to rotate about each of its
principal axes, while a suspended mass connected through
a string and pulley provided the driving torque.
The angular acceleration was determined from the
motion of the falling mass, allowing the corresponding
moment of inertia to be calculated using Newton's laws
of translational and rotational motion. The center of
mass of the payload was also determined experimentally
as a function of the adjustable flight-mount angle
$\theta$. When the measurement axis did not pass through
the center of mass, the parallel-axis theorem was
applied to obtain the corresponding center-of-mass
moments of inertia.

A CAD-based digital twin of the payload was subsequently
developed to calculate the principal moments of inertia
as functions of the flight-mount angle $\theta$.
By systematically varying $\theta$, the evolution
of $I_{xx}$, $I_{yy}$, and $I_{zz}$ was mapped,
allowing identification of configurations in which
the nominal spin axis becomes the intermediate
principal axis and is therefore rotationally unstable.
The experimentally measured inertia values showed
excellent agreement with the CAD-based predictions,
providing independent validation of the digital
twin model.

Representative values are presented in
Table \ref{tab:inertia_values}. The validated inertia
map predicts stable rotation at $\theta=0^\circ$,
intermediate-axis instability at
$\theta=15^\circ$, $30^\circ$, and $45^\circ$,
and a return to stable rotation at
$\theta=60^\circ$ or $\theta=75^\circ$ . These results demonstrate
that varying only the flight-mount angle changes
the ordering of the principal moments of inertia,
allowing the same payload to transition between
stable and unstable rotational regimes.
This provides a quantitative basis for selecting
flight configurations and experimentally testing
the geometry-controlled onset and suppression
of the Dzhanibekov effect.

\begin{figure}[h]
\centering
\includegraphics[width=0.8\linewidth]
{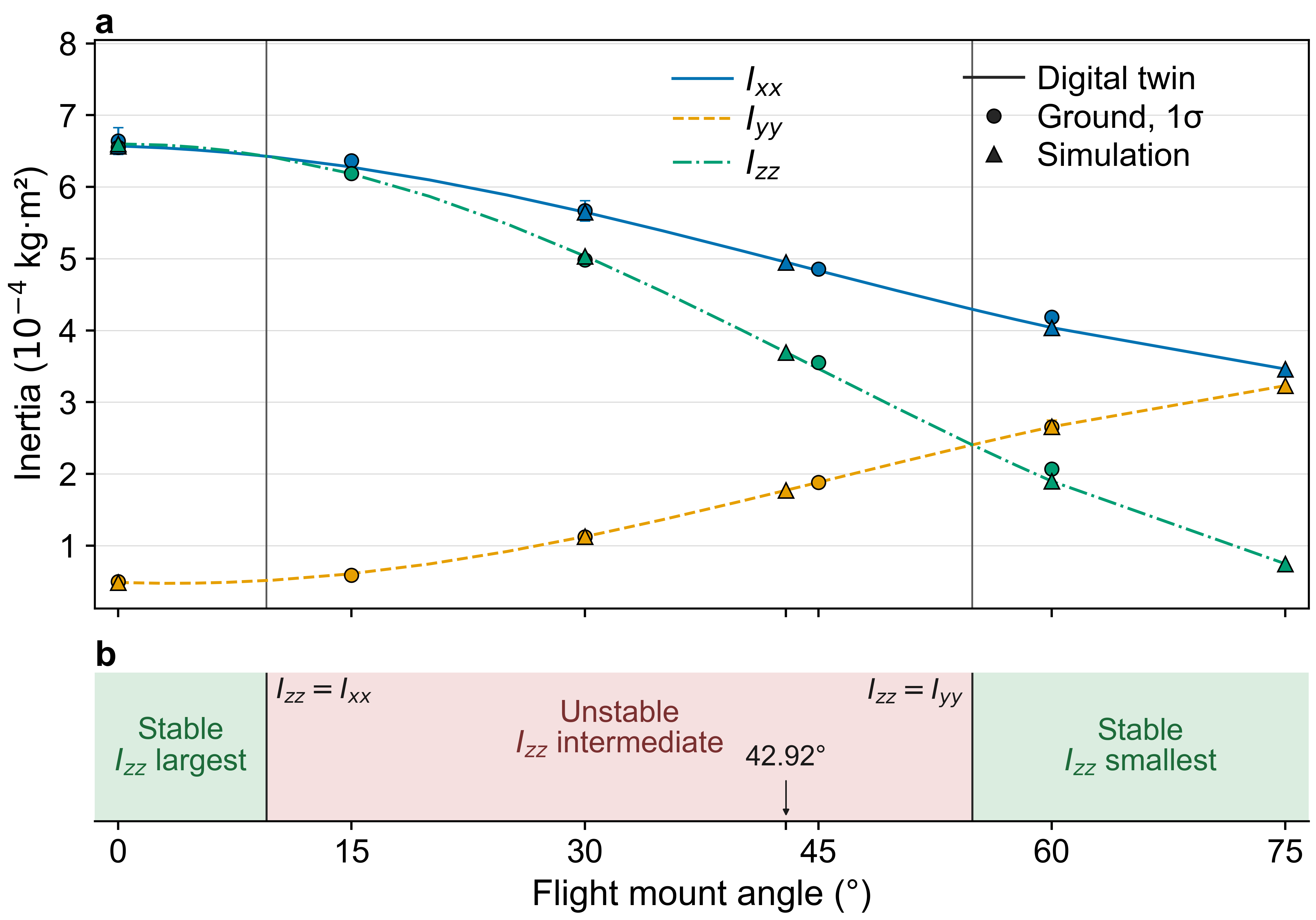}
\caption{
\textbf{Geometry-controlled onset and suppression of
intermediate-axis instability.}
CAD-predicted and experimentally measured principal moments
of inertia, $I_{xx}$, $I_{yy}$, and $I_{zz}$, as functions
of the flight-mount angle $\theta$. The shaded regions
indicate the predicted stability of rotation about the
body-fixed $z$ axis. At small angles, $I_{zz}$ is the
largest principal moment, corresponding to stable rotation.
Near $\theta\simeq 9.5^\circ$, $I_{zz}$ crosses $I_{xx}$
and becomes the intermediate principal moment, marking
the onset of rotational instability. Near
$\theta\simeq 54.5^\circ$, $I_{zz}$ crosses $I_{yy}$
and becomes the smallest principal moment, restoring
rotational stability. The ground-based measurements
agree with the CAD predictions, while the selected
flight configuration ($\theta=42.9^\circ$) lies within
the predicted instability window. These results
demonstrate that intermediate-axis flipping can be
enabled or suppressed by adjusting the payload geometry.
}
\label{fig:inertia_stability_window}
\end{figure}

\begin{table}[h]
\centering
\caption{\textbf{Digital-twin principal moments of inertia calculated numerically as a function of arm angle.} Values are listed in units of $10^{-5}\,\mathrm{kg\,m^2}$. The final column indicates whether the body $z$ axis is predicted to be the intermediate principal axis.}
\label{tab:inertia_values}
\begin{tabular}{ccccc}
\toprule
Arm angle $\theta$ & $I_{xx}$ & $I_{yy}$ & $I_{zz}$ & $z$ axis intermediate?\\
\midrule
$0^\circ$  & 65.71 & 4.86  & 66.00 & No\\
$15^\circ$ & 62.89 & 6.01  & 62.03 & Yes\\
$30^\circ$ & 56.02 & 11.68 & 49.49 & Yes\\
$45^\circ$ & 48.90 & 18.29 & 35.77 & Yes\\
$60^\circ$ & 41.51 & 25.46 & 21.21 & No\\
$75^\circ$ & 34.62 & 32.30 & 74.76 & No\\
\bottomrule
\end{tabular}
\end{table}

The stability map in Fig. \ref{fig:inertia_stability_window} is the essential bridge between object design and flight dynamics. As $\theta$ is increased, $I_{zz}$ decreases strongly, $I_{yy}$ increases, and $I_{xx}$ decreases more gradually. At small angles, $I_{zz}$ is not intermediate, so rotation about the body $z$ axis should not produce Dzhanibekov flips. After the first crossing, $I_{zz}$ becomes intermediate and the same object is predicted to flip when released with nominal $z$-axis spin. After the second crossing, $I_{zz}$ becomes the smallest principal moment and the same nominal spin becomes stable again. The predicted behavior therefore follows a stable--unstable--stable sequence,
\begin{equation}
\text{no flip}\quad\longrightarrow\quad\text{flip}\quad\longrightarrow\quad\text{no flip},
\label{eq:stable_unstable_stable_sequence}
\end{equation}
controlled by a single geometric parameter.

The experimental inertia measurements follow the CAD-predicted curves within the plotted uncertainties, validating the digital twin as a predictive model for the flight configuration. The $42.9^\circ$ case lies well inside the unstable region, where $I_{zz}$ is intermediate. The experimental configurations for the parabolic flights were therefore selected based on the inertia map, which predicted intermediate-axis flipping upon release in microgravity within a specific range of the flight-mount angle $\theta$. Conversely, configurations outside the shaded instability window provide stable-spin controls. This map is the headline experimental result: by tuning the shape of the payload, one can determine in advance whether a freely rotating object will remain in stable spin or undergo large-amplitude inversion.

\subsection{Microgravity experiments}
\begin{figure}[h]
\centering
\includegraphics[width=0.98\linewidth]{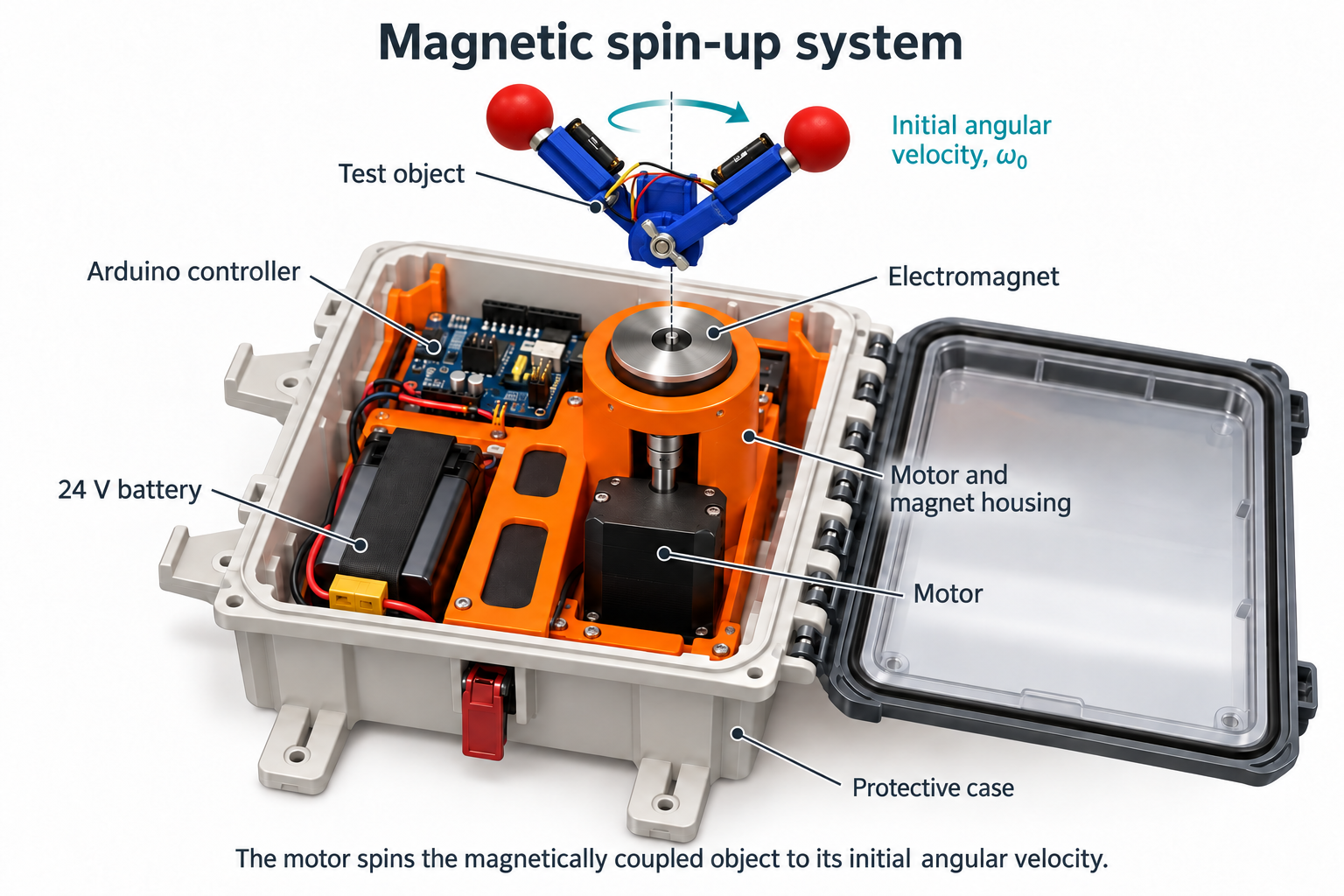}
\caption{
The payload mounted on an a rotating electromagnet.
}
\label{fig:payload_mount}
\end{figure}

The payload was tested during parabolic flight aboard a Falcon 20 aircraft. During the microgravity segment of a parabola, the aircraft approximately follows a ballistic trajectory, reducing the effective acceleration experienced by the payload. If $v$ is the aircraft speed and $R$ is the local radius of curvature of the flight path, the effective acceleration experienced by an object onboard normal to the flight path is approximately
\begin{equation}
g_{\mathrm{eff}} \simeq g_{local}-\frac{v^2}{R},
\label{eq:effective_g}
\end{equation}
where $g_{local}$ is the gravitational acceleration at the elevation of the parabolic flight during the experiment. Microgravity is approached when $v^2/R\simeq g_{local}$. This environment reduces the gravitational torque that would otherwise obscure torque-free Euler dynamics in a ground experiment.

For each trial in the parabolic flight, the payload was spun about the nominal body $z$ axis to approximately $300\,\mathrm{rpm}$, corresponding to
\begin{equation}
\Omega_0\approx 31.4\,\mathrm{rad\,s^{-1}}.
\label{eq:spin_rate}
\end{equation}
The payload was held by an electromagnet prior to release (\cref{fig:payload_mount}). During the microgravity portion of the parabola, the electromagnet was deactivated and the payload separated from the release fixture. Small transverse angular-velocity components introduced during spin-up and release act as perturbations. In stable configurations these perturbations remain bounded; in intermediate-axis configurations they grow and lead to large-amplitude inversion.

\begin{figure}[t]
\centering
\includegraphics[width=0.8\linewidth]{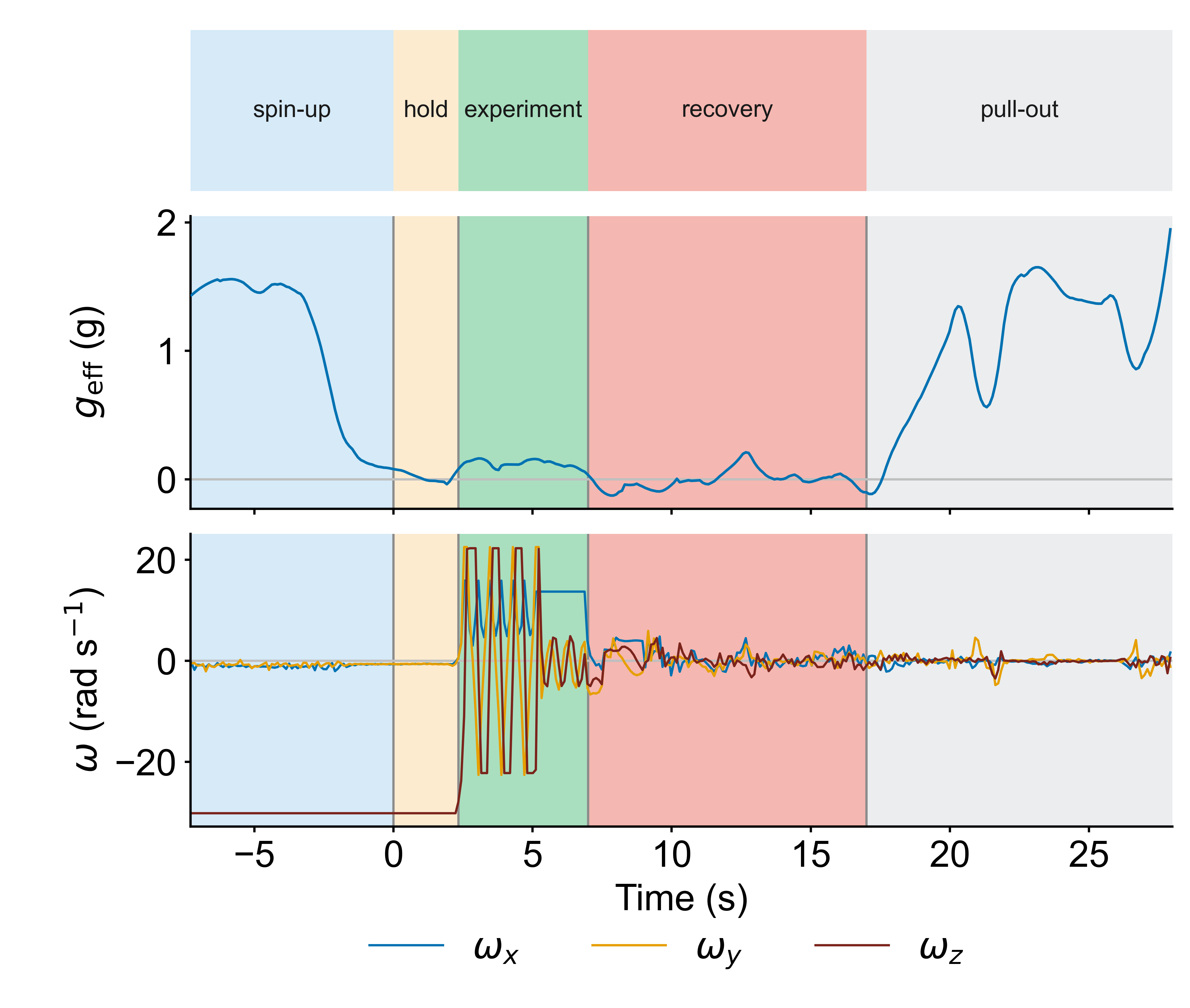}

\caption{
\textbf{Parabolic-flight microgravity profile and experimental
release sequence.}
(A) Measured effective acceleration during the parabola,
showing the transition from elevated acceleration during
pull-up to near-zero acceleration during microgravity.
(B) Three-axis angular velocity recorded by the onboard
inertial measurement unit (IMU), showing the payload
spin-up and release at approximately $t=2.3$ s,
followed by free rotational motion.
The shaded regions identify the spin-up, hold,
experiment, recovery, and pull-out phases.
The payload was released during the microgravity
interval to minimize external torques and allow
observation of its intrinsic rotational dynamics.
}
\label{fig:Parabolic_profile}
\end{figure}


\subsection{Flipping in microgravity observed only within the
predicted instability window}
\begin{figure}[t]
\centering
\includegraphics[width=0.98\linewidth]{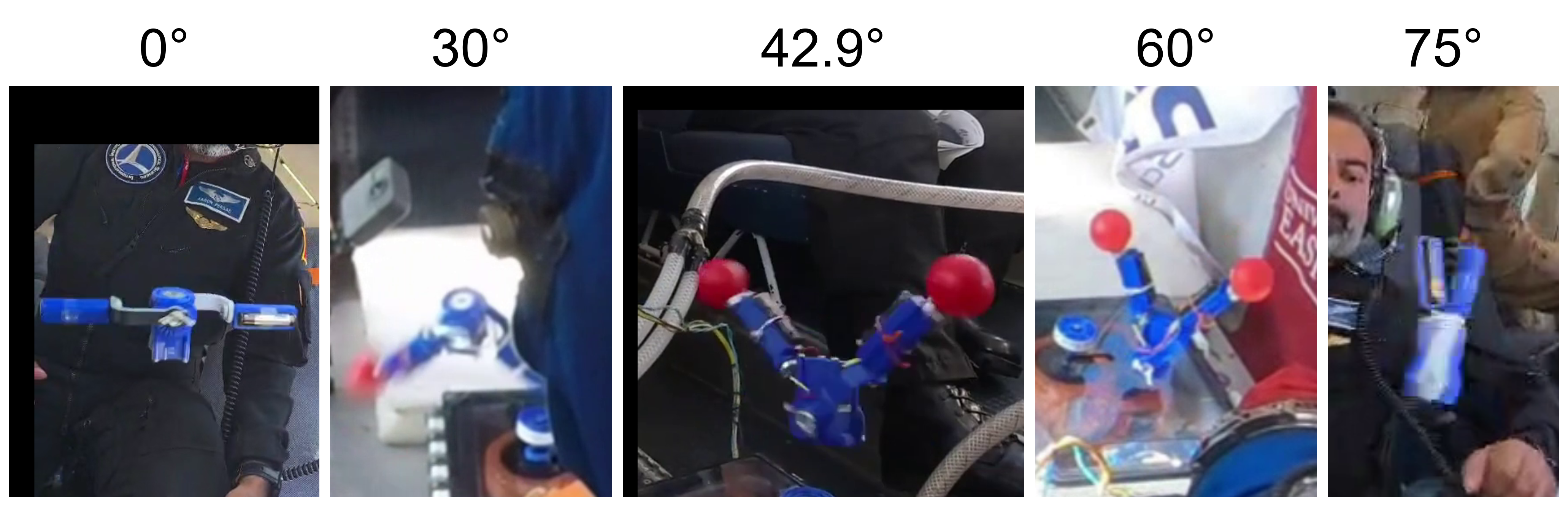}

\caption{
\textbf{Experimental observation of geometry-dependent
rotational stability in microgravity.}
Representative images of the rotating payload during
parabolic-flight experiments at five flight-mount angles:
$\theta = 0^\circ$, $30^\circ$, $42.9^\circ$,
$60^\circ$, and $75^\circ$.
Consistent with the CAD-based digital twin predictions as described in Table \ref{tab:inertia_values},
pronounced intermediate-axis flipping was observed
exclusively at $\theta = 30^\circ$ and $42.9^\circ$,
where the nominal spin axis corresponds to the
intermediate principal moment of inertia.
In contrast, the payload maintained stable rotation
without flipping at $\theta = 0^\circ$, $60^\circ$,
and $75^\circ$, all outside the predicted instability
window. These observations demonstrate that
intermediate-axis instability can be enabled or
suppressed by adjusting the payload geometry.
The corresponding flight recordings are provided
as Supplementary Videos S1--S5.
}
\label{fig:Collage}
\end{figure}

The central experimental prediction is that intermediate-axis
flipping should occur only when the nominal spin axis
corresponds to the intermediate principal moment of inertia.
To test this prediction, the flight-mount angle $\theta$
was systematically varied during the parabolic-flight
experiments, allowing the same payload to be examined in
different rotational stability regimes.

Five configurations were investigated:
$\theta = 0^\circ$, $30^\circ$, $42.9^\circ$,
$60^\circ$, and $75^\circ$. Consistent with the
CAD-based digital twin predictions, pronounced rotational
flipping was observed exclusively at
$\theta = 30^\circ$ and $42.9^\circ$, both of
which lie within the predicted intermediate-axis
instability window. In contrast, the payload exhibited
stable rotation without flipping at
$\theta = 0^\circ$, $60^\circ$, and $75^\circ$,
all of which fall outside the predicted instability
window. Videos documenting the rotational behavior
of the payload in microgravity for all five
flight-mount configurations are provided as
Supplementary Videos S1--S5.

These observations demonstrate that the occurrence of
intermediate-axis flipping is governed by the ordering
of the principal moments of inertia. By changing only
the flight-mount angle, the mass distribution and
corresponding principal moments of inertia were altered,
causing the nominal spin axis to transition between
stable and unstable rotational regimes. The agreement
between the experimentally observed behavior and the
digital twin predictions provides direct evidence
that the Dzhanibekov instability can be enabled
or suppressed through controlled changes in payload
geometry.

Under microgravity conditions, the rotational dynamics of the
payload were quantified using an onboard inertial measurement
unit (IMU) embedded within the payload, which continuously
recorded the three components of angular velocity in the
body-fixed reference frame.

\begin{equation}
\boldsymbol{\omega}_{\mathrm{IMU}}(t)
=
\omega_x(t)\hat{\boldsymbol{x}}
+
\omega_y(t)\hat{\boldsymbol{y}}
+
\omega_z(t)\hat{\boldsymbol{z}},
\label{eq:imu_signal}
\end{equation}

The intermediate-axis instability was characterized
by the growth of transverse angular-velocity components,
followed by large-amplitude nonlinear variations
associated with the reorientation of the payload.
These signatures were observed in the unstable
configurations, whereas the stable configurations
showed no corresponding flipping behavior.

The characteristic inversion interval,
$T_{\mathrm{flip}}$, was determined from successive
orientation-flip events, identified using the
experimental motion records and corresponding
IMU signals. This quantity provides a direct
experimental measure for comparison with the
predictions of nonlinear torque-free Euler dynamics
and the elliptic-integral description of
intermediate-axis flipping.

\section{Theoretical results}

\subsection{Reproduction of the measured angular-velocity trajectory using Euler dynamics}

To determine whether the observed motion is consistent with torque-free rigid-body dynamics, the measured angular velocities were compared with numerical solutions of Euler's equations,
\begin{align}
I_{xx}\dot{\omega}_x &= (I_{yy}-I_{zz})\omega_y\omega_z,\label{eq:euler1}\\
I_{yy}\dot{\omega}_y &= (I_{zz}-I_{xx})\omega_x\omega_z,\label{eq:euler2}\\
I_{zz}\dot{\omega}_z &= (I_{xx}-I_{yy})\omega_x\omega_y.\label{eq:euler3}
\end{align}
The inertial parameters were taken from the digital twin and experimental inertia measurements. Initial angular velocities were obtained from the early post-release IMU signal after excluding the release transient. Agreement between the measured trajectory and Euler simulation indicates that the observed flipping is not merely a release artifact, but the expected nonlinear response of a rigid body in the intermediate-axis regime.

\begin{figure}[h]
\centering
\includegraphics[width=0.98\linewidth]{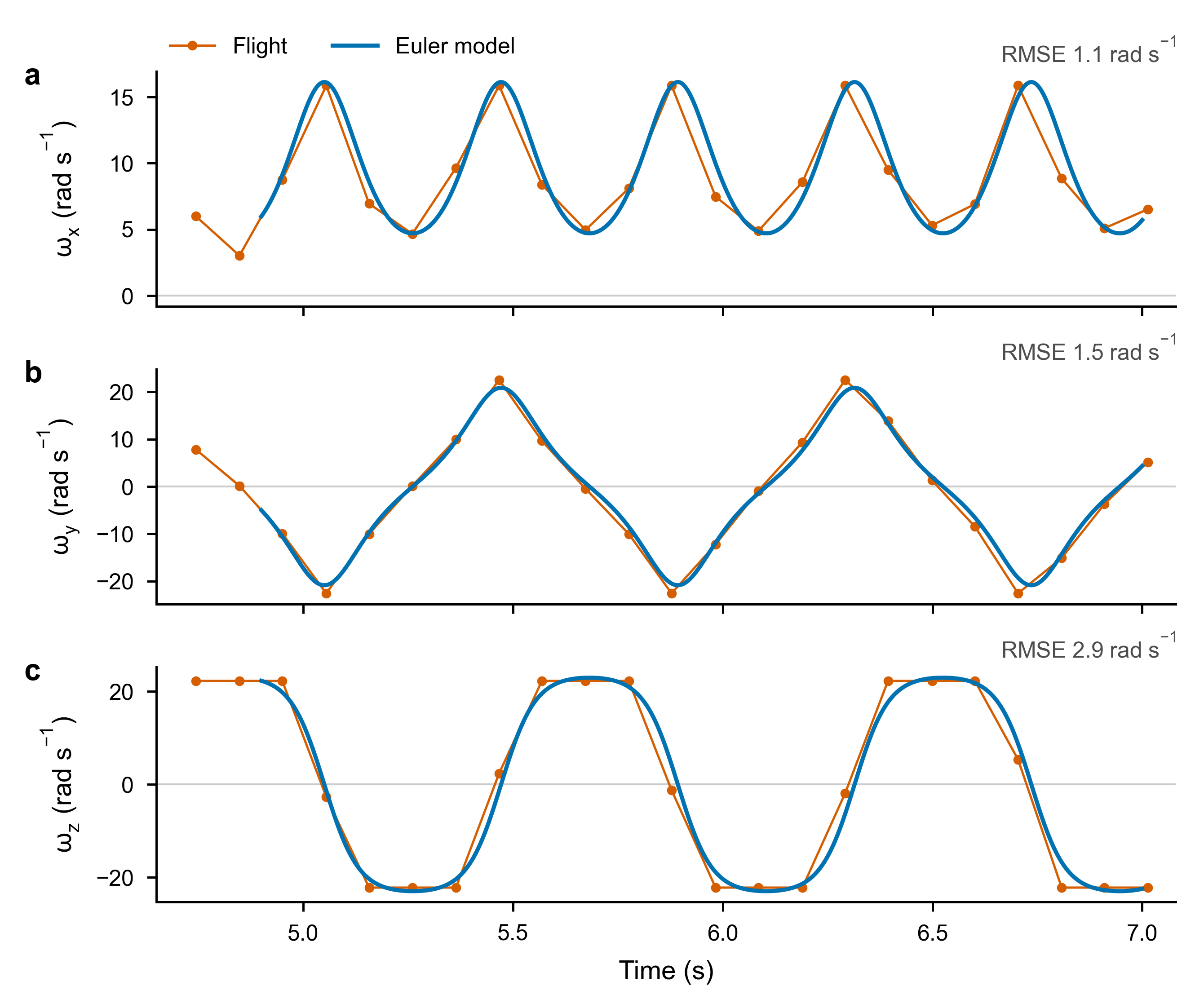}
\caption{This figure shows Euler simulation versus flight IMU data. Overlay measured angular velocities and torque-free Euler simulations for the $42.9^\circ$ configuration are in excellent agreement. 
}
\label{fig:omega_sim_vs_measured}
\end{figure}

\subsection{Poinsot reconstruction validates the conserved geometry}

Torque-free rigid-body motion conserves rotational kinetic energy,
\begin{equation}
E=\frac{1}{2}\left(I_{xx}\omega_x^2+I_{yy}\omega_y^2+I_{zz}\omega_z^2\right),
\label{eq:energy}
\end{equation}
and angular momentum magnitude,
\begin{equation}
L^2=I_{xx}^2\omega_x^2+I_{yy}^2\omega_y^2+I_{zz}^2\omega_z^2.
\label{eq:angular_momentum}
\end{equation}
In angular-momentum coordinates, $L_i=I_{ii}\omega_i$, the constant-angular-momentum surface is a sphere,
\begin{equation}
L_x^2+L_y^2+L_z^2=L^2,
\label{eq:momentum_sphere}
\end{equation}
and the constant-energy surface is an ellipsoid,
\begin{equation}
\frac{L_x^2}{2I_{xx}}+\frac{L_y^2}{2I_{yy}}+\frac{L_z^2}{2I_{zz}}=E.
\label{eq:energy_ellipsoid}
\end{equation}
The body-frame trajectory lies on the intersection of these two surfaces \citep{Peterson2021}. The intermediate-axis state is a saddle of the constrained energy surface, and the separatrix passing near this saddle organizes the flipping trajectory.

The IMU data allow a direct reconstruction of this geometry. The measured angular velocities are converted to angular momentum components using
\begin{equation}
L_i(t)=I_{ii}\omega_i(t).
\label{eq:Li_from_imu}
\end{equation}
Conservation residuals are defined as
\begin{equation}
\delta_L(t)=\frac{|L^2(t)-L^2(0)|}{L^2(0)},
\label{eq:delta_L}
\end{equation}
and
\begin{equation}
\delta_E(t)=\frac{|E(t)-E(0)|}{E(0)}.
\label{eq:delta_E}
\end{equation}
These residuals quantify deviations from ideal torque-free motion due to release impulses, residual accelerations, aerodynamic drag, sensor bias, or unmodeled inertia uncertainty.

\begin{figure}[t]
\centering
\includegraphics[width=0.9\linewidth]{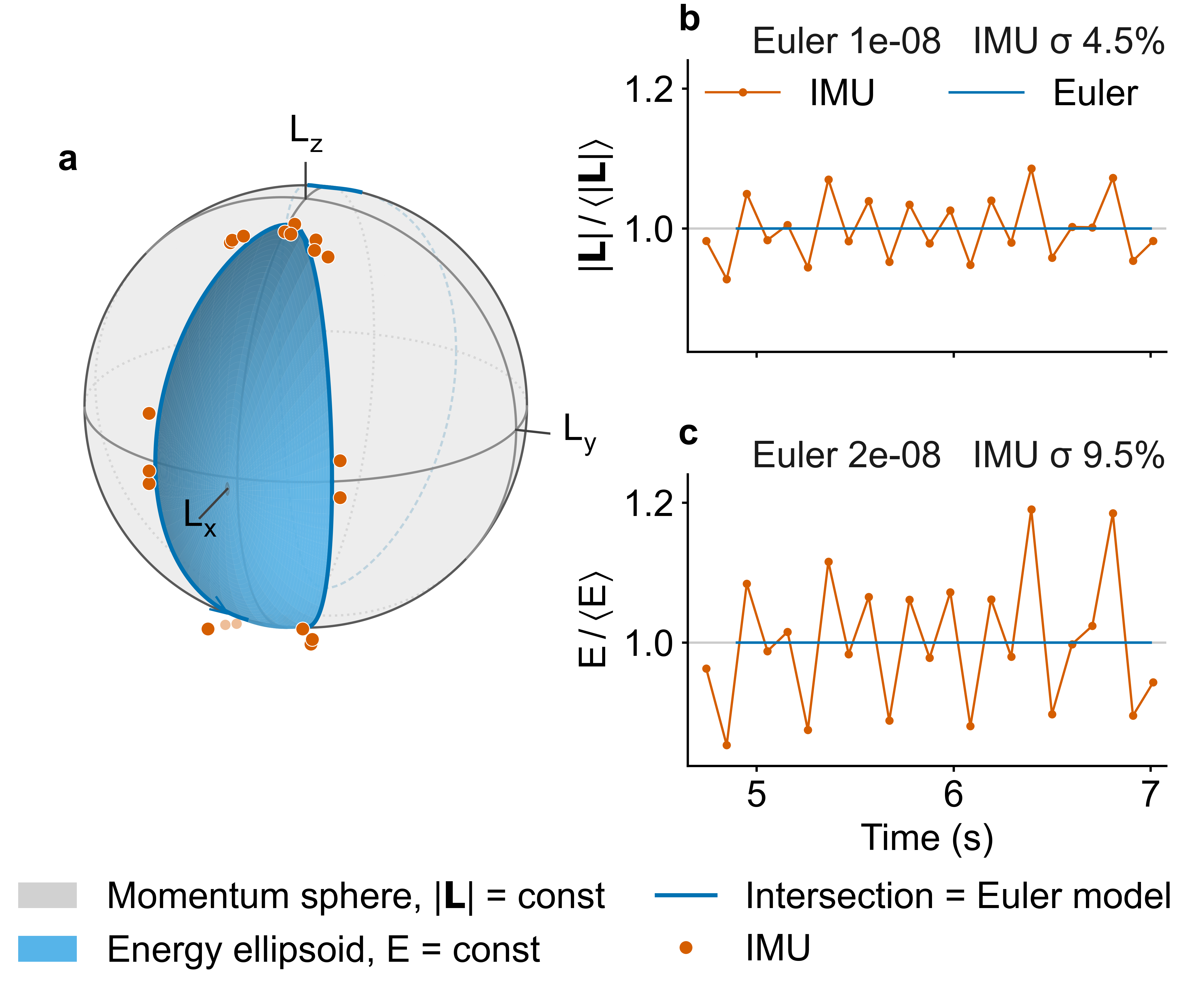}
\caption{
\textbf{Poinsot reconstruction.} Angular-momentum sphere, rotational-energy ellipsoid, separatrix, Euler-simulated trajectory, and IMU-reconstructed trajectory for the unstable configuration. $\delta_L(t)$ and $\delta_E(t)$ quantify the quality of torque-free motion.}
\label{fig:Poinsot}
\end{figure}


\subsection{Separatrix theory predicts the nonlinear flip time}

Linear stability analysis gives the local growth rate near the intermediate-axis rotation,
\begin{equation}
\lambda=\Omega\sqrt{\frac{(I_{zz}-I_{yy})(I_{yy}-I_{xx})}{I_{xx}I_{zz}}},
\label{eq:linear_growth_rate}
\end{equation}
where $I_{xx}<I_{yy}<I_{zz}$ and $\Omega=L/I_{yy}$ is the nominal intermediate-axis spin rate. This expression gives the local exponential growth of perturbations but not the full nonlinear inversion time.

The nonlinear flip time is obtained by reducing Euler's equations using the conserved energy and angular momentum (Eq.\ref{eq:angular_momentum}, and Eq. \ref{eq:momentum_sphere}). For the flipping branch, the experimentally observable half-period can be written as
\begin{equation}
T_{\mathrm{flip}}=\frac{2K(k)}{\nu},
\label{eq:Tflip}
\end{equation}
where
\begin{equation}
K(k)=\int_0^{\pi/2}\frac{\dd\phi}{\sqrt{1-k^2\sin^2\phi}}
\label{eq:complete_elliptic}
\end{equation}
is the complete elliptic integral of the first kind. The parameters $k$ and $\nu$ are determined by the inertia tensor, rotational energy, and angular momentum. 


The separatrix corresponding to pure intermediate-axis rotation occurs at
\begin{equation}
E_{\mathrm{sep}}=\frac{L^2}{2I_{yy}}.
\label{eq:Esep}
\end{equation}
As the trajectory approaches this separatrix, $k\rightarrow1$ and
\begin{equation}
K(k)\sim\frac{1}{2}\ln\left(\frac{16}{1-k^2}\right).
\label{eq:K_asymptotic}
\end{equation}
Therefore the flip time diverges logarithmically near the separatrix. This explains the characteristic dwell-and-flip behavior: the body can remain close to an apparently steady intermediate-axis rotation for an extended time before undergoing rapid inversion.

\begin{figure}[t]
\centering
\includegraphics[width=0.9\linewidth]{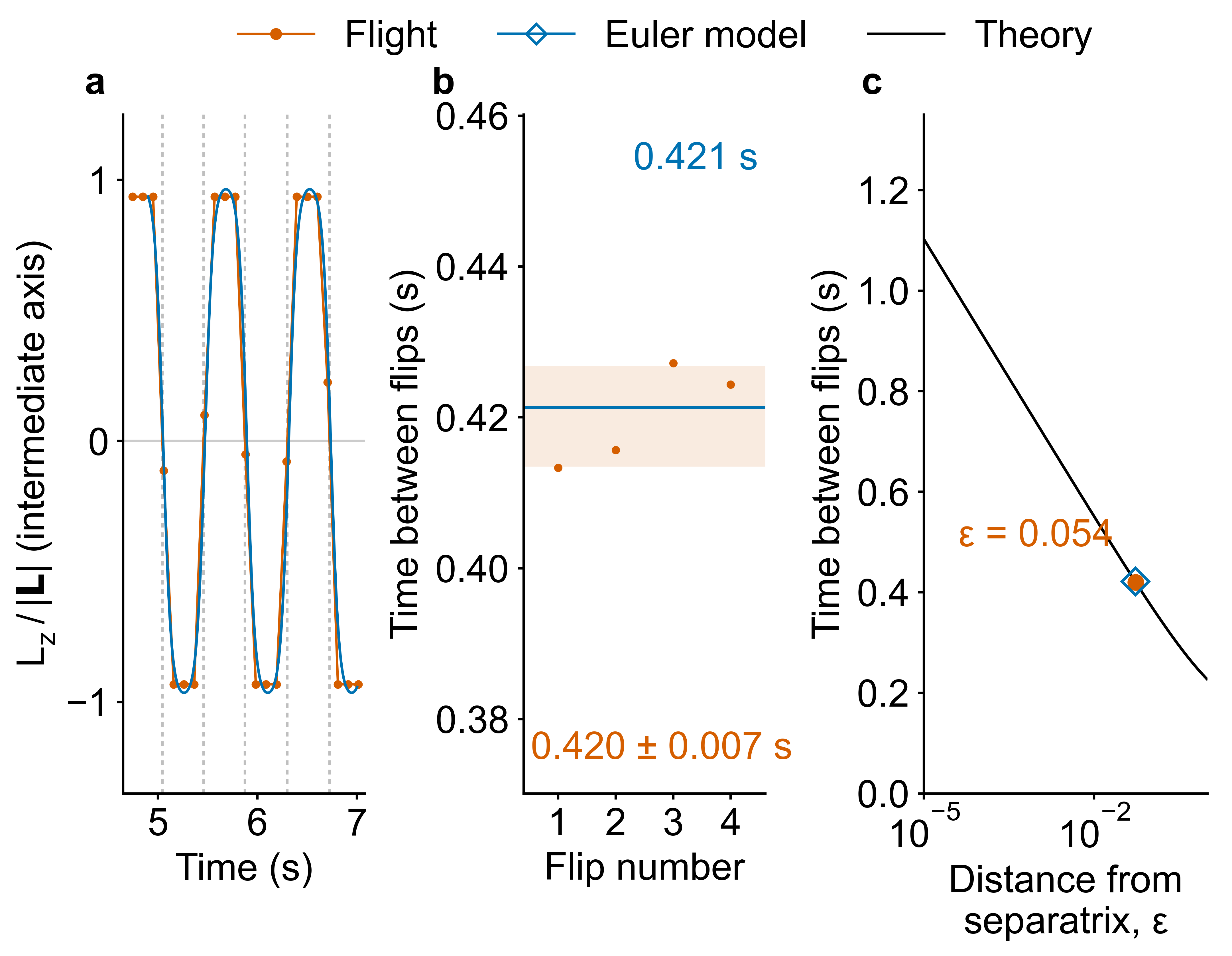}
\caption{
\textbf{Separatrix-controlled flip time.} Measured and simulated normalized flip time $T_{\mathrm{flip}}\Omega$ versus distance from the separatrix $\epsilon=|E-E_{\mathrm{sep}}|/E_{\mathrm{sep}}$.}
\label{fig:Fliptime}
\end{figure}


\subsection{Implications for variable-inertia attitude dynamics}

The experiment shows that the same free-flying body can be moved between stable and unstable attitude regimes by changing its inertia tensor. This result is directly relevant to microgravity systems in which the mass distribution changes during operation, including deployable spacecraft, free-flying robots, articulated payloads, spin-stabilized platforms, and tumbling debris. In these systems, attitude stability can be altered not only by external torques but also by internal geometry. A variable-inertia framework therefore provides a route both to avoiding unwanted tumbling and to deliberately exploiting instability for reorientation.



\section{Discussion}

The present study demonstrates that the onset and suppression
of intermediate-axis instability can be predictably controlled
by changing the mass distribution of a freely rotating rigid
body in microgravity. Although the Dzhanibekov effect is a
well-established consequence of classical rigid-body dynamics,
its experimental demonstration has largely involved objects
with fixed moments of inertia. Here, we move beyond observing
the instability itself and establish a direct experimental
connection between tunable payload geometry, the ordering of
the principal moments of inertia, and the resulting rotational
stability. By combining independently validated inertia
measurements, a CAD-based digital twin, parabolic-flight
experiments, and nonlinear rigid-body theory, we demonstrate
that the same payload can transition between stable and
unstable rotational regimes through a single adjustable
geometric parameter.

The most significant experimental finding is the agreement
between the stability boundaries predicted by the digital
twin and the rotational behavior observed in microgravity.
As the flight-mount angle $\theta$ increases, the principal
moments of inertia evolve such that the nominal spin axis,
corresponding to $I_{zz}$, changes from the largest to the
intermediate and subsequently to the smallest principal
moment. The digital twin predicts an instability window
between approximately $\theta=9.5^\circ$ and
$\theta=54.5^\circ$. Experiments conducted at five distinct
configurations confirmed the predicted stability pattern:
pronounced flipping occurred at $\theta=30^\circ$ and
$42.9^\circ$, whereas no flipping was observed at
$\theta=0^\circ$, $60^\circ$, and $75^\circ$. Thus,
the same payload exhibited a stable--unstable--stable
sequence as its geometry was varied, in agreement with
the independently determined inertia map.

This result provides stronger experimental evidence for
geometry-controlled rotational instability than a
demonstration of flipping at a single configuration.
The stable configurations serve as experimental controls,
showing that spinning and releasing the payload in
microgravity does not necessarily produce inversion.
Instead, the appearance of flipping is determined by
whether the nominal spin axis corresponds to the
intermediate principal moment of inertia. Importantly,
the stability classifications were established from the
independently validated inertia model before comparison
with the observed flight dynamics. The agreement across
all five tested configurations therefore demonstrates
the predictive capability of the digital twin rather
than a retrospective interpretation of the observed
motion.

The microgravity environment plays an essential role in
establishing this connection. On the ground, support
forces, bearing friction, gravitational torques about
fixed pivots, and other mechanical constraints can
complicate the interpretation of rigid-body motion.
During parabolic flight, the payload was released into
a near-free-fall environment, allowing its rotational
dynamics to approximate torque-free motion over the
available observation interval. Although residual
accelerations, aerodynamic drag, and release
perturbations cannot be completely eliminated, the
microgravity environment permits a substantially
cleaner comparison with Euler's equations than
experiments involving mechanically constrained rotation.
The stable configurations also help distinguish
inertia-driven instability from effects associated
with the release mechanism.

Beyond the qualitative observation of flipping,
the measured body-frame angular velocities provide
a quantitative test of nonlinear rigid-body dynamics.
For the unstable $\theta=42.9^\circ$ configuration,
numerical integration of Euler's torque-free equations,
using the independently determined principal moments
of inertia and measured initial angular velocities,
reproduces the principal features of the experimental
IMU trajectories. Agreement across the three angular
velocity components demonstrates that the observed
large-amplitude motion is consistent with the nonlinear
coupling predicted by Euler's equations. This
comparison is particularly important because the
Dzhanibekov effect is frequently presented as a
visual phenomenon, whereas the present measurements
connect the observed inversions directly to the
underlying angular-velocity dynamics.

A complementary interpretation emerges from Poinsot's
geometric construction. In the body-fixed angular
momentum space, conservation of angular momentum
magnitude defines a sphere, while conservation of
rotational kinetic energy defines an ellipsoid.
The torque-free trajectory is constrained to their
intersection. Rotation about the intermediate
principal axis corresponds to a saddle point of
the energy surface constrained to constant angular
momentum magnitude, and the associated separatrix
organizes the large-amplitude rotational motion.
The IMU-based reconstruction provides an experimental
connection between this classical geometric
description and the measured microgravity dynamics.
The approximate conservation of angular momentum
magnitude and rotational energy further supports
the interpretation of the observed motion as
nearly torque-free rotation.

The experiment also provides an opportunity to
examine the nonlinear time scale associated with
intermediate-axis flipping. Linear stability
analysis predicts exponential growth of small
transverse perturbations, but does not describe
the complete inversion. The nonlinear solution
of Euler's equations, expressed in terms of
elliptic integrals, accounts for the finite-amplitude
motion and predicts logarithmic slowing as the
trajectory approaches the intermediate-axis
separatrix. For the primary unstable configuration,
the experimentally measured interval between
successive inversions was
$T_{\mathrm{flip}}=0.420\pm0.007\,\mathrm{s}$,
in close agreement with the corresponding
model prediction of approximately
$0.421\,\mathrm{s}$. This agreement connects the
observed inversion time to the independently
characterized inertia tensor and the nonlinear
dynamics of the rotating payload. While the
present measurements are consistent with the
elliptic-integral description, a systematic
experimental verification of the predicted
logarithmic divergence would require measurements
over a wider range of initial conditions and
distances from the separatrix.

These findings have implications beyond the
specific experimental payload. In microgravity
systems, the mass distribution of a spacecraft
or free-flying object may change as a consequence
of deploying appendages, repositioning robotic
components, redistributing internal masses, or
reconfiguring mechanical structures. Such changes
can alter the ordering of the principal moments
of inertia and consequently the stability of
rotation about a given body-fixed axis. The
present experiment demonstrates this fundamental
mechanism under controlled conditions, showing
that an initially stable nominal spin axis can
become unstable following a change in geometry,
even without requiring an external destabilizing
torque after release.

This observation is relevant to spacecraft
attitude dynamics, deployable structures,
articulated robotic systems, and rotating
space debris. For spacecraft design, a
geometry-dependent stability map could help
identify configurations that should be avoided
during deployment or free rotation. Conversely,
the same physical principles suggest
opportunities for exploiting controlled
changes in mass distribution to assist
reorientation in appropriately designed
free-flying systems. Although the present
experiment changes the payload geometry
between releases rather than during free
flight, it establishes the underlying
stability relationship required to investigate
such variable-inertia maneuvers.

Several limitations should be considered.
The approximately 17-second microgravity
interval restricts the duration over which
the dynamics can be observed, while residual
accelerations, aerodynamic drag, release
transients, and IMU measurement uncertainties
introduce departures from ideal torque-free
motion. The experimental configurations also
sample discrete values of $\theta$ rather
than resolving the stability boundaries
continuously. Consequently, the critical
angles are predicted by the validated digital
twin, while the flight measurements confirm
the corresponding stability classifications
at the selected configurations. Future
experiments with a denser sampling of
flight-mount angles near the predicted
boundaries, improved optical attitude
tracking, and longer-duration microgravity
conditions could provide more precise
measurements of the onset of instability
and its nonlinear temporal evolution.

Taken together, these results establish a
quantitative experimental framework linking
payload geometry, principal moments of
inertia, rotational stability, and
nonlinear flipping dynamics in microgravity.
The central advance is the ability to predict,
before release, whether a particular
configuration of the same rotating object
will remain stable or undergo intermediate-axis
inversion. By demonstrating the predicted
appearance and disappearance of flipping
and connecting the observed motion to Euler
dynamics, Poinsot geometry, and nonlinear
flip-time predictions, this study extends
the Dzhanibekov effect from a classical
illustration of rigid-body instability
to an experimentally testable framework
for understanding and predicting the
attitude dynamics of variable-geometry
systems in microgravity.

\section{Methods}

\subsection{Payload design}

The experimental payload consisted of a 3D-printed PLA
central body with two adjustable arms, batteries, mechanical
fasteners, magnets, retaining rings, table-tennis balls,
and an embedded inertial measurement unit (IMU).
The arm angle $\theta$ was measured from the positive
body-fixed $y$ axis toward the positive $z$ axis and was
adjusted between trials to vary the mass distribution.

The total mass of the fully assembled configuration,
determined from the individual component masses,
was 191.91~g. Component masses, dimensions, materials,
and effective densities are provided in
Table~\ref{tab:components}.

\begin{table}[h]
\centering
\caption{Payload component specifications and effective
CAD densities. Masses and dimensions are reported
per component.}
\label{tab:components}
\resizebox{\textwidth}{!}{%
\begin{tabular}{lcrllc}
\toprule
Component & Qty. & Mass (g) & Dimensions (mm)
& Material & $\rho$ (g/cm$^3$) \\
\midrule

Arm & 2 & 3.82
& $75.0\times31.3\times19.1$ & PLA & 0.714 \\

Battery & 2 & 11.70
& $\varnothing10.5\times44.5$ & Alkaline & 3.450 \\

Battery holder assembly & 2 & 2.45
& $50.0\times12.5\times11.0$ & Plastic/steel & 1.990 \\

Battery holder & 2 & 8.17
& $57.9\times25.2\times20.2$ & PLA & 0.633 \\

Main bolt & 2 & 18.99
& M8, length 45.6 & Steel & 6.270 \\

Nut & 2 & 4.57
& M8, thickness 6.5 & Steel & 6.880 \\

IMU assembly & 1 & 8.36
& $20.2\times20.0\times19.6$ & Electronics & 2.520 \\

Central body & 1 & 36.94
& $50.0\times38.1\times38.1$ & PLA & 1.060 \\

Body cover & 1 & 2.44
& $31.5\times31.5\times7.6$ & PLA & 0.671 \\

Small bolt & 4 & 0.80
& M3, length 9.8 & Steel & 7.320 \\

Threaded rod & 1 & 18.07
& $\varnothing7.9\times54.0$ & Steel & 6.880 \\

Fixed base magnet & 1 & 11.90
& $\varnothing15.2\times8.7$ & NdFeB & 7.54 \\

Stack magnet & 2 & 1.725
& $\varnothing6.35\times6.36$ & NdFeB & 8.56 \\

Retaining ring & 2 & 1.425
& OD 16, ID 8, thickness 6 & Plastic & 1.57 \\

Table-tennis ball & 2 & 2.65
& $\varnothing39.8$ & Plastic & 0.0803 \\

\midrule
\textbf{Total assembly} & & \textbf{191.91}
& & & \\
\bottomrule
\end{tabular}%
}
\end{table}

The onboard data acquisition system consisted of a TinyDuino
processor, a nine-axis IMU TinyShield, a microSD TinyShield,
and a USB TinyShield manufactured by TinyCircuits.
The electronic components are listed in
Table~\ref{tab:imu}.

\begin{table}[h]
\centering
\caption{Components of the onboard data acquisition system.}
\label{tab:imu}
\resizebox{\textwidth}{!}{%
\begin{tabular}{llccr}
\toprule
Component & Part number & Qty.
& Dimensions (mm) & Unit mass (g) \\
\midrule

TinyDuino processor & ASM2001 Rev.~9 & 1
& $20.0\times20.2\times7.5$ & 1.341 \\

9-axis IMU TinyShield & ASD2511 Rev.~5 & 1
& $20.0\times20.0\times5.1$ & 2.084 \\

microSD TinyShield & ASD2201 Rev.~2 & 1
& $20.0\times20.0\times5.1$ & 2.501 \\

USB TinyShield & ASD2101 Rev.~6 & 1
& $20.0\times20.0\times5.5$ & 1.869 \\

Stack spacer & -- & 12
& $3.0\times3.0\times3.5$ & 0.051 \\

\midrule
\textbf{Total assembly} & & & & \textbf{8.407} \\
\bottomrule
\end{tabular}%
}
\end{table}


\subsection{Digital-twin inertia model}

A CAD-based digital twin was constructed in Autodesk
Fusion (version 2706.1.17, x86\_64) using experimentally
measured component dimensions and masses. Component
dimensions were measured using a dial caliper with a
tolerance of $\pm 0.01$~mm. Each component was assigned
an effective density,

\begin{equation}
    \rho_i = \frac{m_i}{V_i},
\end{equation}

where $m_i$ is the measured component mass and $V_i$ is
its CAD volume. The assigned densities are reported
in Table~\ref{tab:components}.

The table-tennis balls were modeled as homogeneous solid
spheres rather than hollow spherical shells. An effective
density of 0.0803~g/cm$^3$ was assigned to preserve the
measured mass of 2.65~g per ball. This approximation
preserves the component mass and center-of-mass location
but neglects the hollow-shell mass distribution.

The fully assembled digital twin had a total mass of
191.88~g, differing from the sum of the measured
component masses by 0.03~g (0.016\%).

The body-fixed $y$ and $z$ axes were horizontal and
vertical, respectively, in the reference configuration,
with the nominal spin axis corresponding to the $z$ axis.
The inertia tensor was calculated about the center of
mass of each assembly using the CAD mass-property
calculations and expressed in the body-fixed coordinate
system:

\begin{equation}
\mathbf{I} =
\begin{bmatrix}
I_{xx} & I_{xy} & I_{xz} \\
I_{xy} & I_{yy} & I_{yz} \\
I_{xz} & I_{yz} & I_{zz}
\end{bmatrix}.
\end{equation}

The principal moments and corresponding principal axes
were determined by eigendecomposition of the inertia
tensor. The digital-twin predictions were independently
validated using falling-mass inertia measurements at
arm angles of $0^\circ$, $15^\circ$, $30^\circ$,
$45^\circ$, and $60^\circ$.


\subsection{Falling-mass inertia measurement}

Moments of inertia were independently measured using a falling-mass pulley method. A known mass $m$ descended while unwinding a string from a pulley of radius $r$ attached to the test object. The falling mass had acceleration $a$, and the payload angular acceleration was $\alpha$, with
\begin{equation}
a=r\alpha.
\end{equation}
The string tension was
\begin{equation}
T=m(g-a),
\end{equation}
and the torque on the payload was $\tau=Tr$. Thus,
\begin{equation}
I\alpha=m(g-a)r.
\end{equation}
Using $a=r\alpha$, the moment of inertia about the measurement axis is
\begin{equation}
I=\frac{mgr}{\alpha}-mr^2.
\label{eq:falling_mass_inertia}
\end{equation}
When the measurement axis did not pass through the center of mass, the parallel-axis theorem was applied:
\begin{equation}
I_{\mathrm{CM}}=I_{\mathrm{meas}}-Md^2,
\end{equation}
where $M$ is the payload mass and $d$ is the perpendicular distance between the measurement axis and the corresponding center-of-mass axis.


\subsection{Center-of-mass measurement}

The center of mass was determined by static equilibrium. The payload was supported at two points separated by a known distance $D$, with support reactions $F_A$ and $F_B$. Force balance gives
\begin{equation}
F_A+F_B=Mg.
\end{equation}
Taking moments about support $A$ gives
\begin{equation}
F_BD=Mg\,x_{\mathrm{CM}},
\end{equation}
so that
\begin{equation}
x_{\mathrm{CM}}=\frac{F_B}{Mg}D.
\end{equation}
This measurement was repeated as needed for the different arm-angle configurations.


\subsection{Parabolic flight and release protocol}

\begin{figure}[h]
    \centering
    \includegraphics[width=100mm]{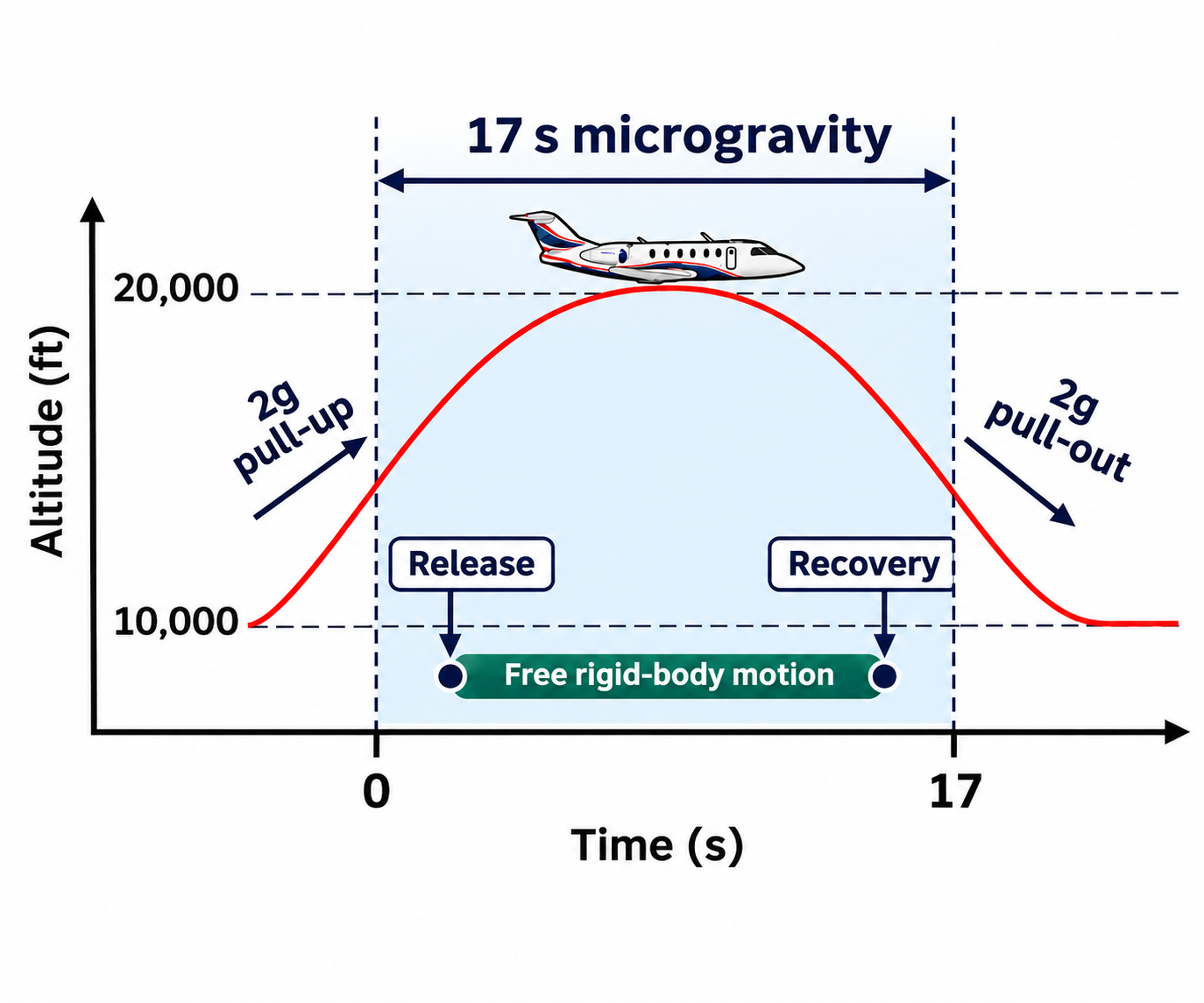}
    \caption{\textbf{Parabolic-flight profile and
    experimental release sequence.}
    The aircraft enters an approximately 17~s
    microgravity interval between the pull-up and
    pull-out phases. Payload release and recovery
    occur within this interval, allowing unconstrained
    rigid-body rotation between these events.
    The trajectory and event positions are schematic.}
    \label{fig:parabolic_flight}
\end{figure}

The payload was flown aboard a Falcon 20 aircraft during parabolic flight as shown in Fig. \ref{fig:parabolic_flight}. Between the pull-up and pull-out phases, the aircraft provides an approximately 17~s microgravity window during which the payload is released, allowed to undergo unconstrained rigid-body rotation, and subsequently recovered. Prior to each release, the payload was spun about the nominal body $z$ axis to approximately $300\,\mathrm{rpm}$. The payload was held by an electromagnet and released during the microgravity interval by deactivating the electromagnet.


\subsection{IMU calibration and angular-velocity processing}

The onboard IMU recorded three-axis angular velocity
with a median sampling interval of 103~ms, corresponding
to a sampling frequency of approximately 9.71~Hz.
The angular-velocity measurements were aligned with
the body-fixed coordinate system and processed using
the corresponding 24~fps flight video.

The processed angular velocities were linearly
interpolated onto a 1~ms time grid for comparison
with the numerical simulations. The quantitative
analysis of the $42.92^\circ$ configuration was
performed over the interval from 4.9 to 7.0~s.


\subsection{Euler-equation simulation}

Torque-free rotational dynamics were simulated using
the complete center-of-mass inertia tensor obtained
from the CAD-based digital twin for each flight
configuration. All off-diagonal inertia tensor
components were retained.

The CAD assembly files corresponding to the nominal
flight configurations are listed in Table~\ref{tab:cad}.
The primary unstable flight configuration had an
arm angle of $42.92^\circ$.

\begin{table}[h]
\centering
\caption{CAD assembly files for the nominal flight
configurations.}
\label{tab:cad}
\begin{tabular}{cl}
\toprule
Arm angle $\theta$ & CAD assembly file \\
\midrule

$0^\circ$  & \texttt{flight\_00deg\_noballs.stl} \\
$30^\circ$ & \texttt{flight\_30deg\_balls.stl} \\
$45^\circ$ & \texttt{flight\_45deg\_balls.stl} \\
$60^\circ$ & \texttt{flight\_60deg\_balls.stl} \\
$75^\circ$ & \texttt{flight\_75deg\_noballs.stl} \\

\bottomrule
\end{tabular}
\end{table}

The rotational dynamics were calculated from Euler's
torque-free rigid-body equation in the body-fixed
coordinate system,

\begin{equation}
\mathbf{I}\dot{\boldsymbol{\omega}}
+
\boldsymbol{\omega}
\times
(\mathbf{I}\boldsymbol{\omega})
=
\mathbf{0},
\end{equation}

which was rearranged for numerical integration as

\begin{equation}
\dot{\boldsymbol{\omega}}
=
-\mathbf{I}^{-1}
\left[
\boldsymbol{\omega}
\times
(\mathbf{I}\boldsymbol{\omega})
\right].
\end{equation}

Initial angular velocities were obtained from the
post-release IMU measurements. The diagonal inertia
components and initial angular velocities used in
each simulation are provided in
Table~\ref{tab:inertia}.

\begin{table}[h]
\centering
\caption{CAD-derived center-of-mass inertia components
and initial angular velocities used in the flight
simulations. Off-diagonal inertia tensor components
were retained for all configurations.}
\label{tab:inertia}
\resizebox{\textwidth}{!}{%
\begin{tabular}{ccrrrc}
\toprule
& & \multicolumn{3}{c}{Inertia (g\,mm$^2$)}
& Initial angular velocity (rad/s) \\
\cmidrule(lr){3-5}
$\theta$ & Event & $I_{xx}$ & $I_{yy}$ & $I_{zz}$
& $(\omega_x,\omega_y,\omega_z)$ \\
\midrule

$0^\circ$ & 1
& 657100 & 48641 & 660000
& $(-0.98,\ 1.05,\ 13.82)$ \\
\midrule

$30^\circ$ & 1
& 564800 & 112800 & 503500
& $(3.90,\ -4.12,\ -28.98)$ \\
\midrule

$42.92^\circ$ & 1
& 495100 & 177200 & 369500
& $(5.83,\ -5.17,\ 22.31)$ \\
\midrule

$60^\circ$ & 1
& 403900 & 265700 & 189800
& $(0.52,\ 0.37,\ -14.51)$ \\
& 2 & & &
& $(0.52,\ 0.34,\ -19.90)$ \\
\midrule

$75^\circ$ & 1
& 346200 & 323000 & 74761
& $(-0.56,\ 3.22,\ 35.97)$ \\
& 2 & & &
& $(-1.27,\ 1.80,\ 35.79)$ \\
& 3 & & &
& $(0.01,\ 0.16,\ 54.34)$ \\
& 4 & & &
& $(0.01,\ 0.17,\ 57.83)$ \\
\bottomrule
\end{tabular}%
}
\end{table}

Euler's equations were numerically integrated using
SciPy (version 1.15.1) with the adaptive Dormand--Prince
Runge--Kutta solver (RK45). Relative and absolute
tolerances were set to $10^{-9}$ and $10^{-9}$~rad/s,
respectively. Solutions were evaluated at the experimental
IMU sampling times or at a uniform frequency of 120~Hz
($\Delta t = 8.33$~ms) for video comparisons. All
simulations retained the complete inertia tensor,
including off-diagonal components.

The simulated angular-velocity trajectories were
compared with the corresponding flight IMU measurements
in the body-fixed coordinate system.

\subsection{Flip-period prediction}

The nonlinear flip time was computed from \cref{eq:Tflip}. The angular momentum and rotational energy were obtained from the measured initial angular velocity and the selected inertia tensor. The distance from the separatrix was quantified using
\begin{equation}
\epsilon=\left|\frac{E-E_{\mathrm{sep}}}{E_{\mathrm{sep}}}\right|.
\end{equation}
The predicted divergence of $T_{\mathrm{flip}}\Omega$ as $\epsilon\rightarrow0$ was compared with simulations and, where available, measured flight data.

\section*{Data availability}
The data that support the findings of this study will be made available from the corresponding authors upon reasonable request. 

\section*{Code availability}
Simulation and analysis code used to integrate Euler's equations, reconstruct Poinsot trajectories, and compute flip-period predictions will be made available from the corresponding authors upon reasonable request. 

\section*{Acknowledgements}
The authors gratefully acknowledge support from the NSF HBCU-UP program under Grant No. 2510172, the U.S. Department of Education through the MSEIP Award No. P120A70068, the International Institute for Astronautical Sciences, the National Research Council of Canada, NSF LSAMP, and the UMES Honors Program. 

\section*{Author contributions}
B.P., J.D., A.H.P., W.L.W., and K.S.D. contributed to the conception, design, execution, analysis, and interpretation of the experiment. B.P. and J.D. contributed to payload development, data collection, and analysis. A.H.P. and K.S.D. supervised the project and developed the manuscript framework. All authors reviewed and approved the manuscript. 

\section*{Competing interests}
The authors declare no competing interests.

\clearpage
\appendix
\section{Supplementary Information}

\subsection{Supplementary Note 1: Elliptic-integral solution for the flip period}
\label{supp:elliptic_solution}

This note provides the full derivation of \cref{eq:Tflip}
for the nonlinear inversion interval of a freely rotating
rigid body. Starting from Euler's torque-free equations
and the conserved quantities $E$ and $L^2$, two
angular-velocity components are eliminated to obtain a
first-order nonlinear differential equation for the
intermediate-axis component. This equation is reduced to
the standard elliptic-integral form and inverted using
Jacobi elliptic functions. Solutions are obtained on both
sides of the separatrix, and the logarithmic divergence
of the flip period is derived explicitly.

\subsubsection*{S1.1. Governing equations and conserved quantities}

Consider a rigid body with three distinct principal
moments of inertia ordered as
\begin{equation}
    I_1<I_2<I_3.
    \label{eq:SI_inertia_ordering}
\end{equation}

The angular velocity in the body-fixed principal-axis
frame is
\begin{equation}
    \boldsymbol{\omega}
    =
    \omega_1\hat{\mathbf{e}}_1+
    \omega_2\hat{\mathbf{e}}_2+
    \omega_3\hat{\mathbf{e}}_3.
\end{equation}

In the absence of external torque, Euler's equations
are given by
\begin{align}
    I_1\dot{\omega}_1
    &=(I_2-I_3)\omega_2\omega_3,
    \label{eq:SI_Euler1}\\
    I_2\dot{\omega}_2
    &=(I_3-I_1)\omega_3\omega_1,
    \label{eq:SI_Euler2}\\
    I_3\dot{\omega}_3
    &=(I_1-I_2)\omega_1\omega_2.
    \label{eq:SI_Euler3}
\end{align}

The rotational kinetic energy,
\begin{equation}
    E=
    \frac{1}{2}
    \left(
    I_1\omega_1^2+
    I_2\omega_2^2+
    I_3\omega_3^2
    \right),
    \label{eq:SI_energy}
\end{equation}
and the squared magnitude of the angular momentum,
\begin{equation}
    L^2=
    I_1^2\omega_1^2+
    I_2^2\omega_2^2+
    I_3^2\omega_3^2,
    \label{eq:SI_L2}
\end{equation}
remain constant throughout torque-free motion.

The simultaneous conservation of these two quantities
makes the torque-free rigid-body problem integrable.
Although Euler's equations constitute three coupled
nonlinear differential equations, the two conservation
laws allow the dynamics to be reduced to a single
first-order differential equation.

For the present analysis, we focus on the evolution of
$\omega_2$, the angular-velocity component associated
with the intermediate principal axis.

\subsubsection*{S1.2. Elimination of the transverse angular-velocity components}

Multiplying \cref{eq:SI_energy} by $2$ gives
\begin{equation}
    2E=
    I_1\omega_1^2+
    I_2\omega_2^2+
    I_3\omega_3^2.
    \label{eq:SI_energy_expanded}
\end{equation}

Multiplying this equation by $I_3$ yields
\begin{equation}
    2EI_3=
    I_1I_3\omega_1^2+
    I_2I_3\omega_2^2+
    I_3^2\omega_3^2.
\end{equation}

Subtracting \cref{eq:SI_L2} eliminates $\omega_3^2$:
\begin{equation}
    2EI_3-L^2
    =
    I_1(I_3-I_1)\omega_1^2
    +
    I_2(I_3-I_2)\omega_2^2.
\end{equation}

Consequently,
\begin{equation}
    \omega_1^2
    =
    \frac{
    2EI_3-L^2-I_2(I_3-I_2)\omega_2^2
    }{
    I_1(I_3-I_1)
    }.
    \label{eq:SI_omega1_eliminated}
\end{equation}

Similarly, multiplying \cref{eq:SI_energy_expanded}
by $I_1$ and subtracting the resulting equation from
\cref{eq:SI_L2} eliminates $\omega_1^2$:
\begin{equation}
    L^2-2EI_1
    =
    I_2(I_2-I_1)\omega_2^2
    +
    I_3(I_3-I_1)\omega_3^2.
\end{equation}

Therefore,
\begin{equation}
    \omega_3^2
    =
    \frac{
    L^2-2EI_1-I_2(I_2-I_1)\omega_2^2
    }{
    I_3(I_3-I_1)
    }.
    \label{eq:SI_omega3_eliminated}
\end{equation}

For mathematical convenience, define
\begin{align}
    A &=
    \frac{2EI_3-L^2}
    {I_2(I_3-I_2)},
    \label{eq:SI_A}\\
    B &=
    \frac{L^2-2EI_1}
    {I_2(I_2-I_1)}.
    \label{eq:SI_B}
\end{align}

Equations \cref{eq:SI_omega1_eliminated,eq:SI_omega3_eliminated}
can then be rewritten as
\begin{equation}
    \omega_1^2
    =
    \frac{I_2(I_3-I_2)}
    {I_1(I_3-I_1)}
    \left(A-\omega_2^2\right),
    \label{eq:SI_omega1_factored}
\end{equation}
and
\begin{equation}
    \omega_3^2
    =
    \frac{I_2(I_2-I_1)}
    {I_3(I_3-I_1)}
    \left(B-\omega_2^2\right).
    \label{eq:SI_omega3_factored}
\end{equation}

Thus, the transverse angular-velocity components
are determined algebraically by the intermediate-axis
component and the two conserved quantities.

\subsubsection*{S1.3. Reduction to a single nonlinear differential equation}

The second Euler equation can be written as
\begin{equation}
    \dot{\omega}_2
    =
    \frac{I_3-I_1}{I_2}
    \omega_3\omega_1.
\end{equation}

Squaring both sides gives
\begin{equation}
    \dot{\omega}_2^{\,2}
    =
    \frac{(I_3-I_1)^2}{I_2^2}
    \omega_1^2\omega_3^2.
    \label{eq:SI_omega2_squared}
\end{equation}

Substituting
\cref{eq:SI_omega1_factored,eq:SI_omega3_factored}
into \cref{eq:SI_omega2_squared} yields
\begin{align}
    \dot{\omega}_2^{\,2}
    &=
    \frac{(I_3-I_1)^2}{I_2^2}
    \frac{I_2(I_3-I_2)}
    {I_1(I_3-I_1)}
    \frac{I_2(I_2-I_1)}
    {I_3(I_3-I_1)}
    \nonumber\\
    &\qquad\times
    \left(A-\omega_2^2\right)
    \left(B-\omega_2^2\right).
\end{align}

After cancellation of common factors,
\begin{equation}
    \boxed{
    \dot{\omega}_2^{\,2}
    =
    \frac{(I_3-I_2)(I_2-I_1)}
    {I_1I_3}
    \left(A-\omega_2^2\right)
    \left(B-\omega_2^2\right).
    }
    \label{eq:SI_master_equation}
\end{equation}

Introducing the dimensionless inertia-asymmetry
parameter
\begin{equation}
    \eta=
    \sqrt{
    \frac{(I_3-I_2)(I_2-I_1)}
    {I_1I_3}
    },
    \label{eq:SI_eta}
\end{equation}
the reduced equation becomes
\begin{equation}
    \dot{\omega}_2^{\,2}
    =
    \eta^2
    \left(A-\omega_2^2\right)
    \left(B-\omega_2^2\right).
    \label{eq:SI_master_eta}
\end{equation}

This equation contains the complete nonlinear
dynamics of the intermediate-axis angular-velocity
component. Its quartic structure naturally leads
to elliptic functions.

\subsubsection*{S1.4. Energy regimes and the separatrix}

For a prescribed angular momentum magnitude $L$,
the rotational energy satisfies
\begin{equation}
    \frac{L^2}{2I_3}
    \leq E \leq
    \frac{L^2}{2I_1}.
\end{equation}

The energy associated with pure rotation about the
intermediate principal axis is
\begin{equation}
    E_{\mathrm{sep}}
    =
    \frac{L^2}{2I_2}.
    \label{eq:SI_Esep}
\end{equation}

This energy defines the separatrix that divides two
families of torque-free trajectories:

\begin{enumerate}
    \item The high-energy branch,
    \begin{equation}
        \frac{L^2}{2I_2}
        <E<
        \frac{L^2}{2I_1},
        \label{eq:SI_high_energy}
    \end{equation}
    corresponding to trajectories surrounding the
    minimum-inertia principal axis.

    \item The low-energy branch,
    \begin{equation}
        \frac{L^2}{2I_3}
        <E<
        \frac{L^2}{2I_2},
        \label{eq:SI_low_energy}
    \end{equation}
    corresponding to trajectories surrounding the
    maximum-inertia principal axis.
\end{enumerate}

The difference between $A$ and $B$ is
\begin{equation}
    A-B
    =
    \frac{
    (I_3-I_1)(2EI_2-L^2)
    }{
    I_2(I_3-I_2)(I_2-I_1)
    }.
    \label{eq:SI_AB_difference}
\end{equation}

Therefore,
\begin{equation}
    \begin{cases}
        A>B, & E>E_{\mathrm{sep}},\\
        A=B, & E=E_{\mathrm{sep}},\\
        A<B, & E<E_{\mathrm{sep}}.
    \end{cases}
\end{equation}

The equality $A=B$ corresponds to the separatrix.
Both energy branches can exhibit pronounced
intermediate-axis inversions when their trajectories
lie sufficiently close to this separatrix.

\subsubsection*{S1.5. Elliptic solution for the high-energy branch}

We first consider the energy regime
$E>E_{\mathrm{sep}}$, for which $0<B<A$.

Since $\omega_2^2$ cannot exceed $B$, introduce
the normalized variable
\begin{equation}
    u=\frac{\omega_2}{\sqrt{B}}.
    \label{eq:SI_u}
\end{equation}

It follows that
\begin{equation}
    \dot{\omega}_2=\sqrt{B}\dot{u}.
\end{equation}

Substituting into \cref{eq:SI_master_eta} gives
\begin{equation}
    B\dot{u}^{\,2}
    =
    \eta^2
    (A-Bu^2)B(1-u^2).
\end{equation}

Define the elliptic modulus
\begin{equation}
    k_+^2=\frac{B}{A},
    \qquad 0<k_+<1,
    \label{eq:SI_k_plus}
\end{equation}
and the nonlinear angular-frequency parameter
\begin{equation}
    \nu_+=\eta\sqrt{A}.
    \label{eq:SI_nu_plus}
\end{equation}

The normalized equation becomes
\begin{equation}
    \boxed{
    \dot{u}^{\,2}
    =
    \nu_+^2(1-u^2)(1-k_+^2u^2).
    }
    \label{eq:SI_elliptic_equation}
\end{equation}

Choosing an interval over which $u$ increases,
separation of variables gives
\begin{equation}
    \frac{du}
    {\sqrt{(1-u^2)(1-k_+^2u^2)}}
    =
    \nu_+\,dt.
\end{equation}

Let $t_0$ denote a zero crossing of $u$.
Integration gives
\begin{equation}
    \nu_+(t-t_0)
    =
    \int_0^u
    \frac{ds}
    {\sqrt{(1-s^2)(1-k_+^2s^2)}}.
    \label{eq:SI_incomplete_integral}
\end{equation}

With the substitution
\begin{equation}
    s=\sin\phi,
\end{equation}
this becomes
\begin{equation}
    \nu_+(t-t_0)
    =
    \int_0^{\arcsin u}
    \frac{d\phi}
    {\sqrt{1-k_+^2\sin^2\phi}}.
\end{equation}

The integral on the right-hand side is the
incomplete elliptic integral of the first kind,
\begin{equation}
    F(\phi,k)
    =
    \int_0^\phi
    \frac{d\psi}
    {\sqrt{1-k^2\sin^2\psi}}.
\end{equation}

The inverse of this integral defines the Jacobi
elliptic sine function. Therefore,
\begin{equation}
    u(t)=
    \operatorname{sn}
    \left(\nu_+(t-t_0),k_+\right).
\end{equation}

The intermediate-axis angular velocity is thus
\begin{equation}
    \boxed{
    \omega_2(t)
    =
    \sqrt{B}\,
    \operatorname{sn}
    \left(\nu_+(t-t_0),k_+\right).
    }
    \label{eq:SI_omega2_sn}
\end{equation}

Using the Jacobi identities
\begin{align}
    \operatorname{sn}^2(q,k)
    +\operatorname{cn}^2(q,k)&=1,\\
    \operatorname{dn}^2(q,k)
    +k^2\operatorname{sn}^2(q,k)&=1,
\end{align}
the remaining components can be obtained from
\cref{eq:SI_omega1_factored,eq:SI_omega3_factored}.

A representative solution is
\begin{align}
    \omega_1(t)
    &=
    \sqrt{
    \frac{I_2(I_3-I_2)A}
    {I_1(I_3-I_1)}
    }
    \operatorname{dn}(q_+,k_+),
    \label{eq:SI_omega1_dn}\\
    \omega_2(t)
    &=
    \sqrt{B}
    \operatorname{sn}(q_+,k_+),\\
    \omega_3(t)
    &=
    \sqrt{
    \frac{I_2(I_2-I_1)B}
    {I_3(I_3-I_1)}
    }
    \operatorname{cn}(q_+,k_+),
    \label{eq:SI_omega3_cn}
\end{align}
where
\begin{equation}
    q_+=\nu_+(t-t_0).
\end{equation}

The phase $t_0$ and consistent component signs
are determined by the initial angular velocity.
In particular, $\omega_1$ and $\omega_3$ may be
multiplied by signs $s_1$ and $s_3$, respectively,
provided that $\omega_2$ is multiplied by
$s_1s_3$, where $s_1,s_3\in\{-1,+1\}$.

These expressions constitute the exact nonlinear
solution of Euler's torque-free equations for
the high-energy branch.

\subsubsection*{S1.6. Exact recurrence period and inversion interval}

The complete elliptic integral of the first kind
is defined as
\begin{equation}
    K(k)
    =
    \int_0^{\pi/2}
    \frac{d\phi}
    {\sqrt{1-k^2\sin^2\phi}}.
    \label{eq:SI_K}
\end{equation}

The Jacobi functions $\operatorname{sn}$ and
$\operatorname{cn}$ have period $4K(k)$,
whereas $\operatorname{dn}$ has period $2K(k)$.

The full recurrence period of the body-frame
angular-velocity vector is consequently
\begin{equation}
    T_{\mathrm{full}}^{(+)}
    =
    \frac{4K(k_+)}{\nu_+}.
    \label{eq:SI_Tfull_plus}
\end{equation}

The intermediate-axis angular-velocity component
satisfies the half-period symmetry
\begin{equation}
    \omega_2
    \left(
    t+\frac{2K(k_+)}{\nu_+}
    \right)
    =
    -\omega_2(t).
\end{equation}

We therefore define the nonlinear inversion
interval as
\begin{equation}
    \boxed{
    T_{\mathrm{flip}}^{(+)}
    =
    \frac{2K(k_+)}{\nu_+}.
    }
    \label{eq:SI_Tflip_plus}
\end{equation}

Substituting the definitions of $\eta$ and $A$
into \cref{eq:SI_nu_plus} gives
\begin{equation}
    \boxed{
    \nu_+^2
    =
    \frac{
    (I_2-I_1)(2EI_3-L^2)
    }{
    I_1I_2I_3
    }.
    }
    \label{eq:SI_nu_plus_explicit}
\end{equation}

The corresponding elliptic modulus is
\begin{equation}
    \boxed{
    k_+^2
    =
    \frac{
    (L^2-2EI_1)(I_3-I_2)
    }{
    (2EI_3-L^2)(I_2-I_1)
    }.
    }
    \label{eq:SI_k_plus_explicit}
\end{equation}

The exact inversion interval for the high-energy
branch is therefore
\begin{equation}
    \boxed{
    T_{\mathrm{flip}}^{(+)}
    =
    2K(k_+)
    \sqrt{
    \frac{I_1I_2I_3}
    {(I_2-I_1)(2EI_3-L^2)}
    }.
    }
    \label{eq:SI_Tflip_plus_explicit}
\end{equation}

This expression predicts the nonlinear inversion
interval entirely from the principal moments of
inertia and the conserved initial values of
energy and angular momentum.

\subsubsection*{S1.7. Elliptic solution for the low-energy branch}

For the complementary energy regime
$E<E_{\mathrm{sep}}$, the ordering of the roots
is reversed, such that $0<A<B$.

In this case, the appropriate normalization is
\begin{equation}
    u=\frac{\omega_2}{\sqrt{A}}.
\end{equation}

Defining
\begin{equation}
    k_-^2=\frac{A}{B}
\end{equation}
and
\begin{equation}
    \nu_-=\eta\sqrt{B},
\end{equation}
the reduced differential equation again assumes
the standard elliptic form,
\begin{equation}
    \dot{u}^{\,2}
    =
    \nu_-^2(1-u^2)(1-k_-^2u^2).
\end{equation}

A representative solution is
\begin{align}
    \omega_1(t)
    &=
    \sqrt{
    \frac{I_2(I_3-I_2)A}
    {I_1(I_3-I_1)}
    }
    \operatorname{cn}(q_-,k_-),\\
    \omega_2(t)
    &=
    \sqrt{A}
    \operatorname{sn}(q_-,k_-),\\
    \omega_3(t)
    &=
    \sqrt{
    \frac{I_2(I_2-I_1)B}
    {I_3(I_3-I_1)}
    }
    \operatorname{dn}(q_-,k_-),
\end{align}
where
\begin{equation}
    q_-=\nu_-(t-t_0).
\end{equation}

As before, the phase and component signs are
selected to satisfy the initial conditions.

The nonlinear angular-frequency parameter is
\begin{equation}
    \boxed{
    \nu_-^2
    =
    \frac{
    (I_3-I_2)(L^2-2EI_1)
    }{
    I_1I_2I_3
    }.
    }
    \label{eq:SI_nu_minus_explicit}
\end{equation}

The elliptic modulus becomes
\begin{equation}
    \boxed{
    k_-^2
    =
    \frac{
    (2EI_3-L^2)(I_2-I_1)
    }{
    (L^2-2EI_1)(I_3-I_2)
    }.
    }
    \label{eq:SI_k_minus_explicit}
\end{equation}

The inversion interval for the low-energy
branch is consequently
\begin{equation}
    \boxed{
    T_{\mathrm{flip}}^{(-)}
    =
    2K(k_-)
    \sqrt{
    \frac{I_1I_2I_3}
    {(I_3-I_2)(L^2-2EI_1)}
    }.
    }
    \label{eq:SI_Tflip_minus}
\end{equation}

The two energy branches have distinct
angular-velocity trajectories but exhibit
the same leading logarithmic divergence
as the separatrix is approached.

\subsubsection*{S1.8. Unified expression for the nonlinear inversion interval}

The solutions for the two energy branches can
be combined into a single expression.

Define
\begin{equation}
    M=\max(A,B),
    \qquad
    m=\min(A,B).
\end{equation}

The elliptic modulus and nonlinear frequency
are then
\begin{equation}
    k=\sqrt{\frac{m}{M}},
    \qquad
    \nu=\eta\sqrt{M}.
\end{equation}

The intermediate-axis angular velocity is
\begin{equation}
    \omega_2(t)
    =
    \sqrt{m}\,
    \operatorname{sn}
    \left(\nu(t-t_0),k\right),
\end{equation}
up to the appropriate sign convention.

The exact nonlinear inversion interval is
therefore
\begin{equation}
    \boxed{
    T_{\mathrm{flip}}
    =
    \frac{2}{\eta\sqrt{\max(A,B)}}
    K\left(
    \sqrt{
    \frac{\min(A,B)}
    {\max(A,B)}
    }
    \right).
    }
    \label{eq:SI_Tflip_unified}
\end{equation}

This unified expression is particularly
convenient for numerical implementation,
as it does not require a separate analytical
formula for each energy branch.

\subsubsection*{S1.9. Exact dynamics on the separatrix}

At the separatrix,
\begin{equation}
    E=E_{\mathrm{sep}}
    =\frac{L^2}{2I_2}.
\end{equation}

Define the reference intermediate-axis
angular velocity as
\begin{equation}
    \Omega=\frac{L}{I_2}.
    \label{eq:SI_Omega}
\end{equation}

Substitution into the definitions of $A$ and
$B$ gives
\begin{equation}
    A=B=\Omega^2.
\end{equation}

The nonlinear differential equation
\cref{eq:SI_master_eta} reduces to
\begin{equation}
    \dot{\omega}_2^{\,2}
    =
    \eta^2(\Omega^2-\omega_2^2)^2.
\end{equation}

For the increasing heteroclinic branch,
\begin{equation}
    \frac{d\omega_2}
    {\Omega^2-\omega_2^2}
    =
    \eta\,dt.
\end{equation}

Integration gives
\begin{equation}
    \frac{1}{\Omega}
    \operatorname{artanh}
    \left(\frac{\omega_2}{\Omega}\right)
    =
    \eta(t-t_0).
\end{equation}

Consequently,
\begin{equation}
    \boxed{
    \omega_2(t)
    =
    \Omega
    \tanh\left[\lambda(t-t_0)\right],
    }
    \label{eq:SI_separatrix_tanh}
\end{equation}
where
\begin{equation}
    \boxed{
    \lambda
    =
    \Omega\eta
    =
    \Omega
    \sqrt{
    \frac{(I_3-I_2)(I_2-I_1)}
    {I_1I_3}
    }.
    }
    \label{eq:SI_lambda}
\end{equation}

The other two angular-velocity components
are
\begin{align}
    \omega_1(t)
    &=
    \Omega
    \sqrt{
    \frac{I_2(I_3-I_2)}
    {I_1(I_3-I_1)}
    }
    \operatorname{sech}
    \left[\lambda(t-t_0)\right],
    \label{eq:SI_separatrix_omega1}\\
    \omega_3(t)
    &=
    \Omega
    \sqrt{
    \frac{I_2(I_2-I_1)}
    {I_3(I_3-I_1)}
    }
    \operatorname{sech}
    \left[\lambda(t-t_0)\right].
    \label{eq:SI_separatrix_omega3}
\end{align}

These expressions are consistent with the
limiting identities of the Jacobi elliptic
functions,
\begin{align}
    \operatorname{sn}(q,1)&=\tanh q,\\
    \operatorname{cn}(q,1)&=\operatorname{sech}q,\\
    \operatorname{dn}(q,1)&=\operatorname{sech}q.
\end{align}

The separatrix solution connects two unstable
intermediate-axis equilibrium states:
\begin{equation}
    (0,-\Omega,0)
    \longrightarrow
    (0,+\Omega,0).
\end{equation}

This heteroclinic trajectory approaches the
equilibrium states asymptotically as
$t\rightarrow\pm\infty$ and therefore has
no finite recurrence period.

Pure intermediate-axis rotation,
$\boldsymbol{\omega}=(0,\pm\Omega,0)$,
is itself an exact equilibrium solution
and does not spontaneously depart from
that state in the absence of perturbations.

\subsubsection*{S1.10. Logarithmic divergence near the separatrix}

As the rotational energy approaches the
separatrix from either side,
\begin{equation}
    E\rightarrow E_{\mathrm{sep}},
\end{equation}
the two roots $A$ and $B$ approach one another.

Consequently,
\begin{equation}
    k^2\rightarrow1,
    \qquad
    \nu\rightarrow\lambda.
\end{equation}

The complete elliptic integral of the first
kind has the asymptotic expansion
\begin{equation}
    K(k)
    =
    \ln\left(\frac{4}{\sqrt{1-k^2}}\right)
    +
    \mathcal{O}\left[
    (1-k^2)\ln\left(\frac{1}{1-k^2}\right)
    \right].
\end{equation}

Equivalently, to leading order,
\begin{equation}
    K(k)
    \sim
    \frac{1}{2}
    \ln\left(\frac{16}{1-k^2}\right).
    \label{eq:SI_K_asymptotic}
\end{equation}

The inversion interval therefore becomes
\begin{equation}
    \boxed{
    T_{\mathrm{flip}}
    \sim
    \frac{1}{\lambda}
    \ln\left(\frac{16}{1-k^2}\right).
    }
    \label{eq:SI_Tflip_asymptotic_k}
\end{equation}

This result demonstrates that the
intermediate-axis inversion interval
diverges logarithmically as the trajectory
approaches the separatrix.

Physically, the divergence arises because
the body spends increasingly long intervals
near the unstable intermediate-axis
equilibria before undergoing a comparatively
rapid inversion. The characteristic duration
of the rapid transition remains controlled
by $\lambda^{-1}$, whereas the dwell time
increases logarithmically.

\subsubsection*{S1.11. Explicit energy-distance scaling near the separatrix}

To quantify the distance from the
separatrix, introduce the signed
dimensionless energy parameter
\begin{equation}
    \delta
    =
    \frac{E-E_{\mathrm{sep}}}
    {E_{\mathrm{sep}}}
    =
    \frac{2EI_2}{L^2}-1.
    \label{eq:SI_delta}
\end{equation}

Its absolute magnitude,
\begin{equation}
    \epsilon=|\delta|,
    \label{eq:SI_epsilon}
\end{equation}
measures the normalized energy distance
from the separatrix.

Using $\Omega=L/I_2$, the quantities
$A$ and $B$ can be written as
\begin{align}
    A
    &=
    \Omega^2
    \left(
    1+\frac{I_3}{I_3-I_2}\delta
    \right),\\
    B
    &=
    \Omega^2
    \left(
    1-\frac{I_1}{I_2-I_1}\delta
    \right).
\end{align}

For convenience, define the dimensionless
inertia ratios
\begin{equation}
    a=\frac{I_3}{I_3-I_2},
    \qquad
    b=\frac{I_1}{I_2-I_1}.
\end{equation}

For the high-energy branch, $\delta=\epsilon$,
giving
\begin{equation}
    k_+^2
    =
    \frac{1-b\epsilon}
    {1+a\epsilon}.
\end{equation}

It follows that
\begin{equation}
    1-k_+^2
    =
    \frac{(a+b)\epsilon}
    {1+a\epsilon}.
\end{equation}

Similarly, for the low-energy branch,
$\delta=-\epsilon$, yielding
\begin{equation}
    k_-^2
    =
    \frac{1-a\epsilon}
    {1+b\epsilon}
\end{equation}
and
\begin{equation}
    1-k_-^2
    =
    \frac{(a+b)\epsilon}
    {1+b\epsilon}.
\end{equation}

For fixed, distinct principal moments
and sufficiently small $\epsilon$,
both branches therefore satisfy
\begin{equation}
    1-k^2
    =
    (a+b)\epsilon
    +\mathcal{O}(\epsilon^2).
\end{equation}

Define the dimensionless geometric
factor
\begin{equation}
    \mathcal{G}
    =
    a+b
    =
    \frac{
    I_2(I_3-I_1)
    }{
    (I_3-I_2)(I_2-I_1)
    }.
    \label{eq:SI_G}
\end{equation}

The separatrix asymptotic expression
then becomes
\begin{equation}
    \boxed{
    T_{\mathrm{flip}}
    \sim
    \frac{1}{\lambda}
    \ln\left(
    \frac{16}{\mathcal{G}\epsilon}
    \right).
    }
    \label{eq:SI_Tflip_energy_scaling}
\end{equation}

More precisely, for fixed inertia ratios,
\begin{equation}
    \lambda T_{\mathrm{flip}}
    =
    \ln\left(
    \frac{16}{\mathcal{G}\epsilon}
    \right)
    +
    \mathcal{O}\left[
    \epsilon\ln\left(\frac{1}{\epsilon}\right)
    \right].
    \label{eq:SI_asymptotic_error}
\end{equation}

Substituting $\lambda=\Omega\eta$ gives
\begin{equation}
    \boxed{
    \Omega T_{\mathrm{flip}}
    \sim
    \frac{1}{\eta}
    \ln\left(
    \frac{16}{\mathcal{G}\epsilon}
    \right).
    }
    \label{eq:SI_dimensionless_period}
\end{equation}

Unlike a phenomenological logarithmic
scaling, this expression explicitly
determines both the logarithmic slope
and the inertia-dependent prefactor.

The approximation requires
$\max(a,b)\epsilon\ll1$ and is not uniform
when two principal moments approach
degeneracy.

\subsubsection*{S1.12. Connection with linear instability theory}

Linearization of Euler's equations near
pure intermediate-axis rotation,
\begin{equation}
    \omega_1\simeq0,
    \qquad
    \omega_2\simeq\Omega,
    \qquad
    \omega_3\simeq0,
\end{equation}
gives
\begin{equation}
    \ddot{\omega}_1
    =
    \Omega^2
    \frac{
    (I_3-I_2)(I_2-I_1)
    }{
    I_1I_3
    }
    \omega_1.
\end{equation}

The corresponding exponential growth
rate is
\begin{equation}
    \lambda=\Omega\eta.
\end{equation}

The characteristic instability-growth
timescale is consequently
\begin{equation}
    \tau_{\mathrm{inst}}
    =
    \frac{1}{\lambda}
    =
    \frac{1}{\Omega}
    \sqrt{
    \frac{I_1I_3}
    {(I_3-I_2)(I_2-I_1)}
    }.
    \label{eq:SI_instability_timescale}
\end{equation}

If the principal moments are comparable
in magnitude, and their differences are
characterized by a common scale
$\Delta I$, then
\begin{equation}
    \eta\sim\frac{\Delta I}{I},
\end{equation}
such that
\begin{equation}
    \tau_{\mathrm{inst}}
    \sim
    \frac{1}{\Omega}
    \frac{I}{\Delta I}.
\end{equation}

Thus, weak inertia anisotropy produces
slow exponential growth of transverse
perturbations.

The exact nonlinear solution establishes
the relationship between this local
instability timescale and the global
inversion interval:
\begin{equation}
    \boxed{
    T_{\mathrm{flip}}
    \sim
    \tau_{\mathrm{inst}}
    \ln\left(
    \frac{16}{\mathcal{G}\epsilon}
    \right).
    }
    \label{eq:SI_linear_nonlinear_connection}
\end{equation}

The inversion interval therefore
contains two physically distinct
contributions:

\begin{enumerate}
    \item An intrinsic instability
    timescale, determined by the spin
    rate and the principal moments of
    inertia.

    \item A logarithmic factor determined
    by the proximity of the initial
    rotational state to the separatrix.
\end{enumerate}

The linear instability rate alone is
therefore insufficient to determine
the nonlinear inversion frequency.

\subsubsection*{S1.13. Universal dimensionless scaling}

The exact inversion interval is
\begin{equation}
    T_{\mathrm{flip}}
    =
    \frac{2K(k)}{\nu}.
\end{equation}

Consequently, a dimensionless recurrence
measure can be introduced as
\begin{equation}
    \mathcal{T}
    =
    \nu T_{\mathrm{flip}}.
\end{equation}

The exact solution gives
\begin{equation}
    \boxed{
    \mathcal{T}=2K(k).
    }
    \label{eq:SI_exact_master_curve}
\end{equation}

This expression defines an exact master
curve for torque-free triaxial rigid
bodies. The effects of absolute inertia
scale and angular-velocity scale are
absorbed into $\nu$, while the remaining
dependence on inertia geometry and
initial conditions is contained in $k$.

Near the separatrix, define the scaled
energy distance
\begin{equation}
    \xi=\mathcal{G}\epsilon.
\end{equation}

Since $\nu\rightarrow\lambda$ in this
limit, the corresponding asymptotic
master curve becomes
\begin{equation}
    \boxed{
    \lambda T_{\mathrm{flip}}
    \sim
    \ln\left(\frac{16}{\xi}\right).
    }
    \label{eq:SI_universal_log_scaling}
\end{equation}

The inversion frequency,
\begin{equation}
    f_{\mathrm{flip}}
    =
    \frac{1}{T_{\mathrm{flip}}},
\end{equation}
therefore satisfies
\begin{equation}
    \boxed{
    \frac{f_{\mathrm{flip}}}{\Omega}
    \sim
    \frac{\eta}
    {\ln(16/\xi)}.
    }
    \label{eq:SI_universal_frequency}
\end{equation}

These dimensionless relationships
provide a framework for comparing
rotating bodies with different masses,
dimensions, inertia tensors, spin rates,
and initial perturbations.

The exact scaling can be tested over
the full range of nondegenerate periodic
solutions, whereas the logarithmic
scaling is restricted to trajectories
sufficiently close to the separatrix.

\subsubsection*{S1.14. Application to the tunable-inertia microgravity payload}

In the present experiment, the principal
moments of inertia depend on the
flight-mount angle $\theta$:
\begin{equation}
    I_i=I_i(\theta).
\end{equation}

The payload geometry is held fixed during
each free-rotation interval, so that
Euler's torque-free equations apply
separately to each configuration.

For configurations in which the nominal
body $z$-axis coincides with the
intermediate principal axis, the
ordered principal moments are
identified as
\begin{equation}
    I_1=I_y,
    \qquad
    I_2=I_z,
    \qquad
    I_3=I_x,
\end{equation}
provided that
\begin{equation}
    I_y<I_z<I_x.
\end{equation}

Accordingly, the local instability
growth rate becomes
\begin{equation}
    \boxed{
    \lambda(\theta)
    =
    |\Omega_z|
    \sqrt{
    \frac{
    [I_x(\theta)-I_z(\theta)]
    [I_z(\theta)-I_y(\theta)]
    }{
    I_x(\theta)I_y(\theta)
    }
    }.
    }
    \label{eq:SI_payload_growth_rate}
\end{equation}

This expression connects the
geometry-dependent inertia tensor
directly to the instability timescale.

At the boundaries of the predicted
instability window,
\begin{equation}
    I_z(\theta)=I_x(\theta)
    \quad\text{or}\quad
    I_z(\theta)=I_y(\theta),
\end{equation}
the linear growth rate approaches zero
and the characteristic instability
timescale diverges.

Outside this window, the linearized
growth-rate squared is negative,
corresponding to bounded oscillatory
perturbations rather than exponential
instability.

This result provides a theoretical
explanation for the stable--unstable--stable
sequence obtained by changing the
payload geometry. It also demonstrates
that configurations within the
intermediate-axis window need not
exhibit identical inversion intervals.

In particular, a configuration close
to an inertia-crossing boundary may
possess an unstable nominal spin axis
but fail to exhibit an observable
inversion during the finite microgravity
interval because its instability-growth
timescale is too long.

\subsubsection*{S1.15. Prediction of inversion intervals from flight IMU data}

For each microgravity trial, the initial
angular-velocity vector is obtained
from the calibrated onboard inertial
measurement unit:
\begin{equation}
    \boldsymbol{\omega}_0
    =
    \left(
    \omega_{1,0},
    \omega_{2,0},
    \omega_{3,0}
    \right).
\end{equation}

The measured gyroscope components must
first be transformed into the principal-axis
frame obtained from the inertia tensor.

The initial conserved quantities are
calculated as
\begin{equation}
    E_0
    =
    \frac{1}{2}
    \sum_{i=1}^{3}I_i\omega_{i,0}^2
\end{equation}
and
\begin{equation}
    L_0^2
    =
    \sum_{i=1}^{3}I_i^2\omega_{i,0}^2.
\end{equation}

The corresponding separatrix energy is
\begin{equation}
    E_{\mathrm{sep},0}
    =
    \frac{L_0^2}{2I_2}.
\end{equation}

These quantities determine $A$, $B$,
$k$, and $\nu$, allowing an independent
prediction of the nonlinear inversion
interval through
\begin{equation}
    T_{\mathrm{flip}}^{\mathrm{theory}}
    =
    \frac{2K(k)}{\nu}.
\end{equation}

The experimental inversion interval
can be determined from successive
zero crossings of the intermediate-axis
angular-velocity component:
\begin{equation}
    \omega_2(t_n)=0.
\end{equation}

Successive inversion intervals are
then defined as
\begin{equation}
    T_{\mathrm{flip},n}^{\mathrm{exp}}
    =
    t_{n+1}-t_n.
\end{equation}

For configurations in the intermediate-axis
window, $\omega_2$ corresponds to the
principal-axis component associated
with the nominal body $z$-axis.

The geometric interpretation of this
observable follows from
\begin{equation}
    \cos\beta(t)
    =
    \frac{\mathbf{L}\cdot\hat{\mathbf{e}}_2}
    {L}
    =
    \frac{I_2\omega_2(t)}{L},
    \label{eq:SI_orientation_angle}
\end{equation}
where $\beta$ is the angle between
the intermediate principal axis and
the inertially conserved angular
momentum vector.

Near the separatrix, $\omega_2$
approaches $\pm L/I_2$, corresponding
to nearly parallel and antiparallel
alignment of the intermediate axis
with the angular momentum vector.
Successive sign reversals therefore
describe nearly $180^\circ$
reorientations of this body axis.

Far from the separatrix, however,
sign reversals of $\omega_2$ need not
correspond to large-amplitude
physical inversions. The full
three-dimensional attitude also
need not recur after one complete
angular-velocity period.

For this reason, the experimentally
observed inversions should be
cross-checked using synchronized
video or reconstructed attitude,
where available.

The time from payload release to
the first inversion is additionally
dependent on the initial phase of
the elliptic-function solution
and must be distinguished from
the interval between successive
inversions.

Quantitative agreement between
theory and experiment can be assessed
using the relative period error
\begin{equation}
    \mathcal{E}_T
    =
    \frac{
    \left|
    T_{\mathrm{flip}}^{\mathrm{exp}}
    -
    T_{\mathrm{flip}}^{\mathrm{theory}}
    \right|
    }{
    T_{\mathrm{flip}}^{\mathrm{theory}}
    }.
\end{equation}

Uncertainties in the principal
moments of inertia, initial angular
velocity, IMU calibration, and
release conditions should be
propagated into the theoretical
prediction, particularly near the
separatrix, where the inversion
interval becomes increasingly
sensitive to the initial rotational
state.

\subsubsection*{S1.16. Summary}

The torque-free Euler equations admit
an exact nonlinear solution in terms
of Jacobi elliptic functions.

The principal results of this
derivation are:

\begin{align}
    \eta
    &=
    \sqrt{
    \frac{(I_3-I_2)(I_2-I_1)}
    {I_1I_3}
    },\\[4pt]
    \lambda
    &=
    \Omega\eta,\\[4pt]
    T_{\mathrm{flip}}
    &=
    \frac{2K(k)}{\nu},\\[4pt]
    T_{\mathrm{flip}}
    &\sim
    \frac{1}{\lambda}
    \ln\left(
    \frac{16}{\mathcal{G}\epsilon}
    \right),
    \qquad \epsilon\rightarrow0.
\end{align}

Together, these expressions connect
the local exponential instability,
the global nonlinear inversion
interval, and the separatrix geometry
of torque-free rigid-body motion.

For the tunable microgravity payload,
they establish a direct theoretical
relationship between the adjustable
geometry, the resulting principal
moments of inertia, the initial
rotational state, and the timescale
of intermediate-axis inversion.

The inertia tensor therefore
determines not only whether a
nominal principal-axis rotation
is unstable, but also, together
with the initial angular momentum
and energy, the characteristic
timing of the resulting nonlinear
rotational dynamics.

\subsection{Supplementary Note 2: Stability phase diagram in normalized inertia space}

This note defines normalized inertias
\begin{equation}
\hat{I}_i=\frac{I_i}{I_x+I_y+I_z},\qquad \hat{I}_x+\hat{I}_y+\hat{I}_z=1,
\end{equation}
and shows that the instability region for nominal spin about the body $z$ axis is
\begin{equation}
(\hat{I}_z-\hat{I}_x)(\hat{I}_z-\hat{I}_y)<0.
\end{equation}
The tunable payload traces a path through this simplex as $\theta$ varies.

\subsection{Supplementary Note 3: Error analysis and uncertainty propagation}

This note should include uncertainties in component masses, arm angle, center-of-mass location, falling-mass acceleration, pulley radius, IMU bias, IMU alignment, and release timing. These uncertainties propagate into the principal moments of inertia, simulated Euler trajectories, conservation residuals, and predicted flip period.

\subsection{Supplementary Note 4: Additional simulations and control cases}

This note should include additional Euler simulations for stable and unstable arm-angle configurations, sensitivity to initial transverse perturbations, and comparison between CAD-predicted and experimentally measured inertia tensors.


\clearpage
\begin{thebibliography}{99}
\bibitem{Dipert2023}
Dipert, R. A. Simplified equations for object trajectories in rotating space habitats and space juggling. \textit{npj Microgravity} \textbf{9}(1), 82 (2023).


\bibitem{EulerRigidBody}
Euler, L. Theoria motus corporum solidorum seu rigidorum. A.F. Röse, Rostock and Greifswald (1765).

\bibitem{GoldsteinClassical}
Goldstein, H., Poole, C. P. \& Safko, J. L. \textit{Classical Mechanics}. 3rd edn. Addison-Wesley (2002).

\bibitem{LandauMechanics}
Landau, L. D. \& Lifshitz, E. M. \textit{Mechanics}. 3rd edn. Butterworth-Heinemann (1976).

\bibitem{MarsdenRatiu}
Marsden, J. E. \& Ratiu, T. S. \textit{Introduction to Mechanics and Symmetry}. Springer (1999).

\bibitem{Krishnaswami2020}
Krishnaswami, G. S. \& Vishnu, T. R. The idea of a Lax pair---Part I: Conserved quantities for a dynamical system. \textit{Resonance} \textbf{25}, 1705--1720 (2020).

\bibitem{Mardesic2020}
Marde\v{s}i\'{c}, P., Guillen, G. G., Van Damme, L. \& Sugny, D. Geometric origin of the tennis racket effect. \textit{Physical Review Letters} \textbf{125}, 064301 (2020).




\bibitem{HughesSpacecraft}
Hughes, P. C. \textit{Spacecraft Attitude Dynamics}. Dover Publications (2004).

\bibitem{Ashbaugh1991}
Ashbaugh, M. S., Chicone, C. C. \& Cushman, R. H.
The twisting tennis racket.
\textit{Journal of Dynamics and Differential Equations}
\textbf{3}, 67--85 (1991).

\bibitem{VanDamme2017}
Van Damme, L., Marde\v{s}i\'{c}, P. \& Sugny, D.
The tennis racket effect in a three-dimensional rigid body.
\textit{Physica D: Nonlinear Phenomena}
\textbf{338}, 17--25 (2017).

\bibitem{Peterson2021}
Peterson, C. \& Schwalm, W.
Euler's rigid rotators, Jacobi elliptic functions,
and the Dzhanibekov or tennis racket effect.
\textit{American Journal of Physics}
\textbf{89}, 349--357 (2021).

\bibitem{Ostanin2024}
Ostanin, I. A. \& Sperl, M.
Arbitrary controlled re-orientation of a spinning
body by evolving its tensor of inertia.
\textit{Computer Physics Communications}
\textbf{300}, 109181 (2024).

\bibitem{DeLaTorre2024}
de la Torre, J. A. \& Espa\~{n}ol, P.
Internal dissipation in the Dzhanibekov effect.
\textit{European Journal of Mechanics - A/Solids}
\textbf{106}, 105298 (2024).


\end{thebibliography}
\end{document}